\documentclass{article}
\usepackage{longtable}
\usepackage{graphicx,times}
\begin {document}
%\pagenumbering{arabic}
%\frenchspacing
%parindent 0.0 cm
%\parskip 0.5cm
%\begin{center}
\begin{flushleft}
{\em Article} \\ \vspace{0.2 cm}
{\LARGE
{\bf Electron impact excitation of F-like W~LXVI}
%{\bf Energy levels, oscillator strengths,  and lifetimes  for transitions in W XL}
}\\

\vspace{0.5 cm}

{\bf {Kanti  M.  ~Aggarwal}}\\ 

\vspace*{0.3 cm}

Astrophysics Research Centre, School of Mathematics and Physics, Queen's University Belfast, Belfast BT7 1NN, Northern Ireland, UK\\ 
%\vspace*{0.5 cm} 

e-mail: K.Aggarwal@qub.ac.uk \\

\vspace*{0.20cm}

Received:  13 May 2016; Accepted:  20 July 2016 \\

%\vspace*{1.0 cm}

\vspace{0.5 cm}

%\clearpage

{\bf Abstract:} 
%\begin{abstract}
Electron impact excitation collision strengths are calculated for all transitions among 113 levels of the 2s$^2$2p$^5$, 2s2p$^6$,  2s$^2$2p$^4$3$\ell$, 2s2p$^5$3$\ell$, and 2p$^6$3$\ell$ configurations of F-like W~LXVI. For this purpose Dirac Atomic R-matrix Code (DARC) has been adopted and results are listed over a wide energy range of 1000 to 6000 Ryd. For comparison purpose analogous calculations have also been performed with the Flexible Atomic Code (FAC), and the results obtained are comparable with those from DARC.

%\end{abstract}
\vspace*{0.5cm}
\end{flushleft}

%{\bf PACS} 32.70.Cs, 34.80 Dp
%\clearpage
\hrule

\section{Introduction}

Tungsten (W) is a very important constituent of fusion reactor walls because of its materialistic characteristics. Additionally, it radiates at almost all ionisation stages and therefore atomic data (namely energy levels, radiative rates and collision strengths) are required for many of its ions. These data are required for modelling and diagnosing fusion plasmas for the elemental density, temperature and chemical composition determination. Realising this there have been many experimental and theoretical efforts to obtain atomic data for W ions. Experimentally, only level energies have been determined for many of the W ions, as compiled and assessed by Kramida and Shirai {\cite{ks1}}, and updated by Kramida {\cite{ks2}}. Their recommended energies are also available on the NIST (National Institute of Standards and Technology)  website at {\tt http://www.nist.gov/pml/data/asd.cfm}. However, for most ions the measured energies are available for only a few levels and there are no corresponding measurements for other parameters, i.e. radiative rates (A-values) and collision strengths ($\Omega$). For this reason it becomes necessary to calculate data for the desired atomic parameters.

Theoretically, energy levels and A-values have been determined by many workers {\cite{kbf, mrmp, saf2, saf, pq}} for several W ions. We too have calculated these parameters for a few W ions, such as W~XL {\cite{w40b}}, W~LVIII {\cite{w58b}} and W~LIX to W~LXVI {\cite{w66b}}. Here we focus our attention on F-like W~LXVI. Additionally, for the other W ions we have only reported energy levels and A-values, but here we also calculate results for $\Omega$.  

Earlier calculations have been performed by Sampson {\em et al.} {\cite{zs1}} for F-like  ions with 22 $\le$ Z $\le$ 92, including W~LXVI. They adopted a relativistic atomic structure code to calculate oscillator strengths (f-values) and the {\em distorted-wave} (DW) method for the determination of $\Omega$. However, they presented only limited results for transitions from the  lowest three levels of 2s$^2$2p$^5$ and 2s2p$^6$ to higher excited 110 levels of the  2s$^2$2p$^4$3$\ell$, 2s2p$^5$3$\ell$ and 2p$^6$3$\ell$ configurations. Additionally, they reported values of $\Omega$ at only six energies, approximately in the 140  -- 4400 Ryd range, depending on the transition because their results are in terms of {\em excited} energies. Furthermore, the format adopted by them  is not straightforward to understand and apply their data. Therefore, recently S. Aggarwal {\cite{sa}}  (henceforth to be referred to as SA)  has reported results for energy levels, f-values, A-values, line strengths (S-values),  and lifetimes ($\tau$)   for the lowest 60 levels of the 2s$^2$2p$^5$, 2s2p$^6$ and  2s$^2$2p$^4$3$\ell$ configurations of W~LXVI. For his calculations, he adopted the modified version of the {\sc grasp} (general-purpose relativistic atomic structure package) code,  known as GRASP0. It is a fully relativistic code, is based on the $jj$ coupling scheme, and includes two-body relativistic corrections arising from the Breit interaction and QED (quantum electrodynamics) effects.

Unfortunately, the calculations of SA  (apart from being very limited in range) are found to be erroneous and unreliable, particularly for $\tau$, as recently discussed and demonstrated by us {\cite{w66a}}. Nevertheless, his calculations were extended to all 113 levels of the 2s$^2$2p$^5$, 2s2p$^6$,  2s$^2$2p$^4$3$\ell$, 2s2p$^5$3$\ell$, and 2p$^6$3$\ell$ configurations by Goyal {\em et al.} {\cite{jam}}, who also calculated $\Omega$ but  at only four energies in the 1000 -- 1600 Ryd range for the resonant {\em forbidden} transitions alone. This is because they included only limited range of partial waves with angular momentum $J \le$ 9, {\em insufficient} for the convergence of $\Omega$ for most allowed (and some forbidden) transitions. Therefore, in this work we extend the range of partial waves to $J \le$ 50 and energy to $\le$ 6400 Ryd, so that $\Omega$ can be determined for {\em all} transitions and over a wide energy range.

As by   Goyal {\em et al.} {\cite{jam}}, we adopt the {\em Dirac atomic R-marix code} (DARC) and include the same 113 levels of the  2s$^2$2p$^5$, 2s2p$^6$,  2s$^2$2p$^4$3$\ell$, 2s2p$^5$3$\ell$, and 2p$^6$3$\ell$ configurations. Since results for energy levels and A-values (for four types of transition, namely electric dipole (E1), electric quadrupole (E2), magnetic dipole (M1), and magnetic quadrupole (M2)) have already been reported by us {\cite{w66b}}, we here focus only on the  $\Omega$ data. 

%__________________________________________________________________

\section{Calculations}

The 113 levels of the 2s$^2$2p$^5$, 2s2p$^6$,  2s$^2$2p$^4$3$\ell$, 2s2p$^5$3$\ell$, and 2p$^6$3$\ell$ configurations are listed in Table~1 along with our calculated energies with the {\sc grasp} code. These energies are comparable to those listed by Goyal {\em et al.} {\cite{jam}} and their accuracy has already been discussed in our earlier work {\cite{w66a}}. However, we note  that the energies obtained by Goyal {\em et al.}, with the corresponding {\sc fac} code and listed in their table~1, are {\em incorrect}, particularly for the highest 20 levels. These energies are in clear error, by up to $\sim$60~Ryd. For this reason in Table~1 we have also listed the `correct' energies obtained with the {\sc fac} code. There is a satisfactory agreement  between the {\sc grasp} and {\sc fac} energies, and the discrepancies are below 0.6~Ryd. Similarly,  the orderings are nearly the same between the two calculations, although there are a few minor differences -- see for example, levels 10--12, 17--18 and 43--44.

For consistency, we employ the same DARC code for the calculations of $\Omega$, as by  Goyal {\em et al.} {\cite{jam}}. It is based on the $jj$ coupling scheme, and uses the  Dirac-Coulomb Hamiltonian in the $R$-matrix approach. The $R$-matrix radius adopted for  W~LXVI is 0.66 au, and 25  continuum orbitals have been included for each channel angular momentum in the expansion of the wavefunction, allowing us to compute $\Omega$ up to an energy of  6400 Ryd, i.e. $\sim$5500 Ryd {\em above} the highest threshold. This energy range is  sufficient to calculate values of effective collision strength ($\Upsilon$)  up to T$_e$ = 3 $\times$10$^{8}$~K, well above what may be required for  application to fusion plasmas where high temperatures prevail.  The maximum number of channels for a partial wave is 512, and the corresponding size of the Hamiltonian matrix is 12~816. To obtain convergence of  $\Omega$ for all transitions and at all energies, we have included all partial waves with angular momentum $J \le$ 50. Furthermore,  to account for higher neglected partial waves, we have included a top-up, based on the Coulomb-Bethe approximation {\cite{ab}} for allowed transitions and geometric series for others.

In Table~1 are also listed our calculated $\Omega$ for all resonance transitions at an energy of 1600~Ryd, the highest energy considered by Goyal {\em et al.} {\cite{jam}}. For a ready comparison their corresponding results   are also listed at this energy. Their results are restricted to only a few (not all) \,{\em forbidden} transitions, as they did not calculate $\Omega$ for the allowed ones, because the partial waves range was restricted to $J \le$ 9. As a result of this for a few transitions, such as 1--8 (2s$^2$2p$^2$~$^2$P$^o_{3/2}$ -- 2s$^2$2p$^4$3p~$^2$D$^o_{5/2}$), 1--12 (2s$^2$2p$^2$~$^2$P$^o_{3/2}$ -- 2s$^2$2p$^4$3p~$^4$D$^o_{7/2}$), 1--41 (2s$^2$2p$^2$~$^2$P$^o_{3/2}$ -- 2s$^2$2p$^4$3p~$^2$F$^o_{7/2}$), 1--73 (2s$^2$2p$^2$~$^2$P$^o_{3/2}$ -- 2s2p$^5$3d~$^4$F$^o_{7/2}$), and 1--79 (2s$^2$2p$^2$~$^2$P$^o_{3/2}$ -- 2s2p$^5$3d~$^4$D$^o_{7/2}$), their values of $\Omega$ are {\em underestimated}, by up to 25\%. The reason for this is clearly apparent from Fig. 1 (a, b and c) in which we show the variation of $\Omega$ with $J$, at six energies of 1000, 2000, 3000, 4000, 5000, and 6000~Ryd, and for three transitions, namely 1--3 (2s$^2$2p$^2$~$^2$P$^o_{3/2}$ -- 2s2p$^6$~$^2$S$_{1/2}$), 1--8 (2s$^2$2p$^2$~$^2$P$^o_{3/2}$ -- 2s$^2$2p$^4$3p~$^2$D$^o_{5/2}$) and 2--3 (2s$^2$2p$^2$~$^2$P$^o_{1/2}$ -- 2s2p$^6$~$^2$S$_{1/2}$). The 1--3 and 2--3 transitions are allowed and therefore they converge slowly with $J$. In fact, a slightly longer range of $J$ would have been preferable to have fully converged $\Omega$ for such transitions. However, due to computational limitations it is not possible, and therefore a `top-up' has been included for the additional contributing partial waves, based on the Coulomb-Bethe approximation of Burgess and Sheorey  {\cite{ab}}, as already stated. On the other hand, the 1--8 (see Fig. 1b)  is a forbidden transition but $J \le$ 9 are not sufficient for the convergence of $\Omega$, even at an energy of 1000~Ryd. Therefore, the reported $\Omega$ of Goyal {\em et al.} {\cite{jam}} are clearly underestimated, apart from being limited in the energy range as well as the transitions.

In Table~2 we list our calculated $\Omega$ at 10 energies (1000, 1200, 1400, 1600, 1800, 2000, 3000, 4000, 5000, and 6000~Ryd) for all transitions from the lowest 3 to higher excited levels. The corresponding results for all other remaining transitions may be obtained on request from the author. For the same transitions as listed in Table~2, Sampson {\em et al.} {\cite{zs1}} have also reported $\Omega$, but only in a limited energy range, as already stated. In Fig. 2  we compare our results with them, in the common energy range (below 4000~Ryd)  for the same three transitions (1--3, 1--8 and 2--3), as shown in Fig. 1. The agreement between the two calculations is fully satisfactory and there is no (significant) discrepancy for these three (and other) transitions of W~LXVI.

Finally, in Fig.~3 we compare our results of $\Omega$ from {\sc darc} with the corresponding calculations from {\sc fac} for three transitions, namely 4--16 (2s$^2$2p$^4$3s~$^4$P$_{5/2}$ -- 2s$^2$2p$^4$3d~$^4$D$_{5/2}$), 5--18 (2s$^2$2p$^4$3s~$^2$P$_{3/2}$ -- 2s$^2$2p$^4$3d~$^2$F$_{7/2}$) and 8--10 (2s$^2$2p$^4$3p~$^2$D$^o_{5/2}$ -- 2s$^2$2p$^4$3p~$^4$P$^o_{5/2}$), all of which are forbidden but have comparatively larger magnitude. Both sets of $\Omega$ agree well (within 20\%), which is highly satisfactory and confirms, once again, that there is no (major) discrepancy among the two present calculations with {\sc darc} and {\sc fac}, and the earlier one by  Sampson {\em et al.} {\cite{zs1}}. However, we will like to emphasise here the importance of including a large range of partial waves in order to have accurate determination of $\Omega$. In Fig. 4 we show the variation of $\Omega$ for the same three transitions (as shown in Fig. 3) with a ``top-up" performed at $J$ = 30, 40 and 50. It is a standard practice to top-up the values of $\Omega$ with a geometric series for the forbidden transitions. It is clear from this figure that the results topped-up at  $J$ = 30 can be overestimated by up to a factor of two, in comparison to those at $J$ = 50. Similarly, $\Omega$ obtained with a top-up at $J$ = 40 can be overestimated by up to $\sim$25\%, particularly towards the higher end of the energy range. For this reason there is scope to slightly further improve the accuracy of $\Omega$ at high end of the energy range, by further extending the $J$ range. However, it is not possible due to computational limitations. In conclusion, the accuracy of our calculated $\Omega$ is comparatively higher (within 20\%) for energies up to $\sim$5000~Ryd, but may be  lower at higher energies. However, it is clearly apparent  that the partial waves range ($\le$ 9) included by Goyal   {\em et al.} {\cite{jam}} is grossly inadequate for the determination of accurate $\Omega$ values.

\section{Conclusions}

In this work, we have reported values of $\Omega$ over a wide range of energy, for transitions among the lowest 113 levels of F-like  W LXVI. For the calculations, we have adopted a fully relativistic  scattering code ({\sc darc}), and have included a wide range of partial waves with up to $J$ = 50, generally sufficient for the convergence of $\Omega$ for most transitions and at most energies. Nevertheless, to further improve the accuracy of our calculated $\Omega$, a top-up has also been included to account for the contribution of higher neglected partial waves.  Comparisons of $\Omega$ have been made with the present and earlier {\cite{zs1}} DW calculations and no (major) discrepancy has been noted. Based on these comparisons the accuracy of our calculated $\Omega$ is within 20\%, for most transitions and at energies up to $\sim$5000~Ryd.

A more useful parameter, not considered in the present paper, is {\em effective} collision strength ($\Upsilon$), which is obtained by integrating $\Omega$ values over a wide energy range and with a suitable electron velocity distribution function, generally Maxwellian. However, $\Omega$ in the thresholds region does not vary as smoothly as shown in Figs. 2 and 3, or in Table~2. Because of the Feshbach (closed channel) resonances, it is a highly varying function of energy, and therefore the resonances need to be resolved in a fine energy mesh to accurately determine the values of $\Upsilon$. With the computational resources available with us, such calculations will take several months to conclude, but will be more useful for applications.

%\begin{acknowledgements}
%\section*{Acknowledgment}
 %KMA  is thankful to  AWE Aldermaston for financial support.     
% \end{acknowledgements}

%\newpage

%\end{document}

\newpage
\clearpage
			  
\begin{flushleft}			  
{\bf Table 1.} Comparison of GRASP and FAC energies  (in Ryd) for the levels of  W LXVI, and collision
strengths ($\Omega$) for the resonance transitions at 1600 Ryd. $a{\pm}b \equiv$ $a\times$10$^{{\pm}b}$.
\newline 				  
\end{flushleft} 			  
					  
\begin{tabular}{rllrrrrrrr} \hline	  
Index  &     Configuration          & Level          & GRASP    & FAC      & $\Omega$1     & $\Omega$2        \\
\hline					  
    1 & 2s$^2$2p$^5$		 & $^2$P$^o_{3/2}$   &   0.0000 &   0.0000 &           &              \\
    2 & 2s$^2$2p$^5$		 & $^2$P$^o_{1/2}$   & 102.0364 & 102.1521 & 2.028$-$3 &  1.860$-$3   \\
    3 & 2s2p$^6$		 & $^2$S$  _{1/2}$   & 137.7025 & 137.4448 & 3.990$-$2 &       $ $    \\
    4 & 2s$^2$2p$^4$3s  	 & $^4$P$  _{5/2}$   & 621.0158 & 621.0472 & 4.571$-$4 &       $ $    \\
    5 & 2s$^2$2p$^4$3s  	 & $^2$P$  _{3/2}$   & 621.8299 & 621.8704 & 1.258$-$3 &       $ $    \\
    6 & 2s$^2$2p$^4$3s  	 & $^2$S$  _{1/2}$   & 625.8014 & 625.8038 & 3.349$-$4 &       $ $    \\
    7 & 2s$^2$2p$^4$($^3$P)3p	 & $^4$P$^o_{3/2}$   & 631.1841 & 631.1019 & 3.460$-$4 &  3.010$-$4   \\
    8 & 2s$^2$2p$^4$($^3$P)3p	 & $^2$D$^o_{5/2}$   & 631.3275 & 631.2490 & 9.917$-$4 &  7.550$-$4   \\
    9 & 2s$^2$2p$^4$($^1$S)3p	 & $^2$P$^o_{1/2}$   & 635.7137 & 635.6013 & 2.716$-$4 &  2.160$-$4   \\
   10 & 2s$^2$2p$^4$($^3$P)3p	 & $^4$P$^o_{5/2}$   & 659.8186 & 659.7615 & 4.288$-$4 &  3.770$-$4   \\
   11 & 2s$^2$2p$^4$($^3$P)3p	 & $^2$S$^o_{1/2}$   & 659.8909 & 659.8336 & 1.490$-$4 &  1.350$-$4   \\
   12 & 2s$^2$2p$^4$($^3$P)3p	 & $^4$D$^o_{7/2}$   & 659.8176 & 659.7604 & 4.919$-$4 &  4.080$-$4   \\
   13 & 2s$^2$2p$^4$($^3$P)3p	 & $^4$S$^o_{3/2}$   & 663.4845 & 663.4209 & 4.894$-$3 &  4.810$-$3   \\
   14 & 2s$^2$2p$^4$($^1$S)3p	 & $^2$P$^o_{3/2}$   & 665.3718 & 665.2889 & 8.125$-$3 &  8.160$-$3   \\
   15 & 2s$^2$2p$^4$($^3$P)3d	 & $^4$D$  _{3/2}$   & 670.6545 & 670.5186 & 5.541$-$4 &       $ $    \\
   16 & 2s$^2$2p$^4$($^3$P)3d	 & $^4$D$  _{5/2}$   & 670.8287 & 670.6959 & 1.302$-$3 &       $ $    \\
   17 & 2s$^2$2p$^4$($^3$P)3d	 & $^4$P$  _{1/2}$   & 671.0968 & 670.9598 & 1.531$-$3 &       $ $    \\
   18 & 2s$^2$2p$^4$($^3$P)3d	 & $^2$F$  _{7/2}$   & 670.8850 & 670.7513 & 7.566$-$4 &       $ $    \\
   19 & 2s$^2$2p$^4$($^1$S)3d	 & $^2$D$  _{3/2}$   & 675.3438 & 675.1778 & 4.138$-$4 &       $ $    \\
   20 & 2s$^2$2p$^4$($^3$P)3d	 & $^4$D$  _{7/2}$   & 677.3153 & 677.2001 & 4.780$-$4 &       $ $    \\
   21 & 2s$^2$2p$^4$($^3$P)3d	 & $^4$F$  _{9/2}$   & 677.3630 & 677.2476 & 5.318$-$4 &       $ $    \\
   22 & 2s$^2$2p$^4$($^3$P)3d	 & $^2$P$  _{1/2}$   & 678.4084 & 678.2780 & 3.296$-$3 &       $ $    \\
   23 & 2s$^2$2p$^4$($^3$P)3d	 & $^2$D$  _{5/2}$   & 680.2199 & 680.0693 & 1.682$-$2 &       $ $    \\
   24 & 2s$^2$2p$^4$($^1$D)3d	 & $^2$P$  _{3/2}$   & 680.3181 & 680.1636 & 1.607$-$2 &       $ $    \\
   25 & 2s$^2$2p$^4$($^1$S)3d	 & $^2$D$  _{5/2}$   & 682.9101 & 682.7466 & 1.599$-$2 &       $ $    \\
   26 & 2s$^2$2p$^4$3s  	 & $^4$P$  _{3/2}$   & 723.5924 & 723.7012 & 1.089$-$4 &       $ $    \\
   27 & 2s$^2$2p$^4$3s  	 & $^2$P$  _{1/2}$   & 724.3128 & 724.4307 & 8.681$-$5 &       $ $    \\
   28 & 2s$^2$2p$^4$3s  	 & $^2$D$  _{5/2}$   & 725.0859 & 725.1903 & 2.380$-$4 &       $ $    \\
   29 & 2s$^2$2p$^4$3s  	 & $^2$D$  _{3/2}$   & 725.5494 & 725.6597 & 1.058$-$4 &       $ $    \\
   30 & 2s$^2$2p$^4$($^3$P)3p	 & $^4$P$^o_{1/2}$   & 733.3901 & 733.3812 & 5.235$-$5 &  5.250$-$5   \\
   31 & 2s$^2$2p$^4$($^3$P)3p	 & $^4$D$^o_{3/2}$   & 733.8677 & 733.8586 & 4.172$-$4 &  4.190$-$4   \\
   32 & 2s$^2$2p$^4$($^1$D)3p	 & $^2$F$^o_{5/2}$   & 734.9009 & 734.8889 & 1.591$-$4 &  1.600$-$4   \\
   33 & 2s$^2$2p$^4$($^1$D)3p	 & $^2$P$^o_{3/2}$   & 738.0872 & 738.0779 & 7.557$-$3 &  7.590$-$3   \\
   34 & 2s2p$^5$($^3$P)3s	 & $^4$P$^o_{5/2}$   & 753.9775 & 753.7860 & 1.476$-$4 &  1.490$-$4   \\
   35 & 2s2p$^5$($^3$P)3s	 & $^2$P$^o_{3/2}$   & 756.0908 & 755.8926 & 3.669$-$3 &  3.680$-$3   \\
   36 & 2s2p$^5$($^1$P)3s	 & $^2$P$^o_{1/2}$   & 760.3451 & 760.2761 & 6.399$-$5 &  6.400$-$5   \\
   37 & 2s2p$^5$($^1$P)3s	 & $^2$P$^o_{3/2}$   & 761.3225 & 761.1850 & 5.118$-$3 &  5.140$-$3   \\
\hline	
\end{tabular}
\newpage
\begin{tabular}{rllrrrrrrrrrrr} \hline
Index  &     Configuration                        & Level    & GRASP      &   FAC & $\Omega$1 & $\Omega$2     \\        
\hline  
   38 & 2s$^2$2p$^4$($^3$P)3p	 & $^4$D$^o_{5/2}$   & 762.4810 & 762.4912 & 1.912$-$4 &  1.780$-$4   \\
   39 & 2s$^2$2p$^4$($^3$P)3p	 & $^2$D$^o_{3/2}$   & 762.8417 & 762.8679 & 2.548$-$4 &  2.310$-$4   \\
   40 & 2s$^2$2p$^4$($^3$P)3p	 & $^2$P$^o_{1/2}$   & 763.3247 & 763.2028 & 5.819$-$5 &  5.130$-$5   \\
   41 & 2s$^2$2p$^4$($^1$D)3p	 & $^2$F$^o_{7/2}$   & 763.6274 & 763.6406 & 3.567$-$4 &  2.770$-$4   \\
   42 & 2s2p$^5$($^3$P)3p	 & $^4$P$  _{3/2}$   & 764.6947 & 764.3845 & 6.249$-$4 &       $ $    \\
   43 & 2s$^2$2p$^4$($^1$D)3p	 & $^2$D$^o_{3/2}$   & 764.5814 & 764.4924 & 2.231$-$4 &  2.230$-$4   \\
   44 & 2s2p$^5$($^3$P)3p	 & $^2$D$  _{5/2}$   & 764.7548 & 764.4542 & 2.448$-$3 &       $ $    \\
   45 & 2s$^2$2p$^4$($^1$D)3p	 & $^2$D$^o_{5/2}$   & 764.5917 & 764.6064 & 2.274$-$4 &  2.020$-$4   \\
   46 & 2s$^2$2p$^4$($^1$D)3p	 & $^2$P$^o_{1/2}$   & 768.7542 & 768.7653 & 6.868$-$5 &  6.310$-$5   \\
   47 & 2s2p$^5$($^1$P)3p	 & $^2$P$  _{1/2}$   & 770.7609 & 770.4710 & 1.053$-$3 &       $ $    \\
   48 & 2s2p$^5$($^1$P)3p	 & $^2$D$  _{3/2}$   & 770.9584 & 770.6495 & 1.305$-$3 &       $ $    \\
   49 & 2s$^2$2p$^4$($^3$P)3d	 & $^4$D$  _{1/2}$   & 772.6818 & 772.6419 & 4.247$-$5 &       $ $    \\
   50 & 2s$^2$2p$^4$($^3$P)3d	 & $^4$F$  _{3/2}$   & 773.5498 & 773.5032 & 2.544$-$4 &       $ $    \\
   51 & 2s$^2$2p$^4$($^3$P)3d	 & $^4$F$  _{5/2}$   & 773.9835 & 773.9305 & 2.910$-$3 &       $ $    \\
   52 & 2s$^2$2p$^4$($^1$D)3d	 & $^2$G$  _{7/2}$   & 774.4988 & 774.4520 & 1.712$-$4 &       $ $    \\
   53 & 2s$^2$2p$^4$($^1$D)3d	 & $^2$S$  _{1/2}$   & 776.1136 & 775.9935 & 3.725$-$3 &       $ $    \\
   54 & 2s$^2$2p$^4$($^1$D)3d	 & $^2$F$  _{5/2}$   & 776.1806 & 776.1017 & 9.448$-$3 &       $ $    \\
   55 & 2s$^2$2p$^4$($^1$D)3d	 & $^2$D$  _{3/2}$   & 776.5900 & 776.4849 & 7.625$-$3 &       $ $    \\
   56 & 2s$^2$2p$^4$($^3$P)3d	 & $^4$F$  _{7/2}$   & 779.7429 & 779.7072 & 2.751$-$4 &       $ $    \\
   57 & 2s$^2$2p$^4$($^3$P)3d	 & $^2$F$  _{5/2}$   & 780.7621 & 780.7166 & 2.195$-$4 &       $ $    \\
   58 & 2s$^2$2p$^4$($^3$P)3d	 & $^2$P$  _{3/2}$   & 780.8245 & 780.7799 & 1.418$-$4 &       $ $    \\
   59 & 2s$^2$2p$^4$($^1$D)3d	 & $^2$G$  _{9/2}$   & 781.2456 & 781.2032 & 3.748$-$4 &       $ $    \\
   60 & 2s$^2$2p$^4$($^1$D)3d	 & $^2$D$  _{5/2}$   & 781.9062 & 781.8558 & 2.032$-$4 &       $ $    \\
   61 & 2s$^2$2p$^4$($^1$D)3d	 & $^2$F$  _{7/2}$   & 782.3934 & 782.3387 & 2.880$-$4 &       $ $    \\
   62 & 2s$^2$2p$^4$($^3$P)3d	 & $^2$D$  _{3/2}$   & 784.1105 & 784.0251 & 2.330$-$4 &       $ $    \\
   63 & 2s$^2$2p$^4$($^1$D)3d	 & $^2$P$  _{1/2}$   & 784.5638 & 784.4725 & 1.470$-$4 &       $ $    \\
   64 & 2s2p$^5$($^3$P)3p	 & $^4$D$  _{7/2}$   & 793.0956 & 792.8049 & 1.031$-$4 &       $ $    \\
   65 & 2s2p$^5$($^3$P)3p	 & $^2$P$  _{3/2}$   & 793.6704 & 793.3832 & 5.946$-$4 &       $ $    \\
   66 & 2s2p$^5$($^3$P)3p	 & $^4$P$  _{5/2}$   & 794.1495 & 793.8652 & 5.824$-$4 &       $ $    \\
   67 & 2s2p$^5$($^3$P)3p	 & $^2$P$  _{1/2}$   & 795.7677 & 795.4634 & 3.999$-$4 &       $ $    \\
   68 & 2s2p$^5$($^1$P)3p	 & $^2$D$  _{5/2}$   & 799.7264 & 799.4123 & 6.864$-$4 &       $ $    \\
   69 & 2s2p$^5$($^1$P)3p	 & $^2$P$  _{3/2}$   & 800.2321 & 799.9214 & 2.773$-$4 &       $ $    \\
   70 & 2s2p$^5$($^1$P)3p	 & $^2$S$  _{1/2}$   & 802.5375 & 802.2087 & 3.635$-$5 &       $ $    \\  
   71 & 2s2p$^5$($^3$P)3d	 & $^4$P$^o_{1/2}$   & 803.4369 & 803.0729 & 6.151$-$5 &  5.990$-$5   \\
   72 & 2s2p$^5$($^3$P)3d	 & $^4$P$^o_{3/2}$   & 804.1412 & 803.7745 & 1.978$-$4 &  1.850$-$4   \\
   73 & 2s2p$^5$($^3$P)3d	 & $^4$F$^o_{7/2}$   & 804.3569 & 803.9899 & 1.530$-$3 &  1.260$-$3   \\
   74 & 2s2p$^5$($^3$P)3d	 & $^4$D$^o_{5/2}$   & 804.6674 & 804.2952 & 5.927$-$4 &  5.360$-$4   \\
   75 & 2s2p$^5$($^1$P)3d	 & $^2$P$^o_{1/2}$   & 810.4496 & 810.0549 & 5.359$-$4 &  4.760$-$4   \\
\hline	
\end{tabular}
\newpage
\begin{tabular}{rllrrrrrrrrrrr} \hline
Index  &     Configuration                        & Level    & GRASP      &   FAC & $\Omega$1 & $\Omega$2     \\        
\hline
   76 & 2s2p$^5$($^1$P)3d	 & $^2$F$^o_{5/2}$   & 810.3552 & 809.9675 & 2.897$-$4 &  2.720$-$4   \\
   77 & 2s2p$^5$($^1$P)3d	 & $^2$D$^o_{3/2}$   & 810.6443 & 810.2531 & 2.168$-$4 &  2.020$-$4   \\
   78 & 2s2p$^5$($^3$P)3d	 & $^4$F$^o_{9/2}$   & 810.1949 & 809.8459 & 2.236$-$4 &       $ $    \\
   79 & 2s2p$^5$($^3$P)3d	 & $^4$D$^o_{7/2}$   & 811.7383 & 811.3732 & 1.245$-$3 &  1.040$-$3   \\
   80 & 2s2p$^5$($^3$P)3d	 & $^2$D$^o_{5/2}$   & 811.8379 & 811.4678 & 1.655$-$3 &  1.470$-$3   \\
   81 & 2s2p$^5$($^3$P)3d	 & $^2$D$^o_{3/2}$   & 813.2720 & 812.8860 & 1.900$-$3 &  1.680$-$3   \\
   82 & 2s2p$^5$($^3$P)3d	 & $^2$P$^o_{1/2}$   & 814.7593 & 814.3551 & 5.642$-$4 &  5.030$-$4   \\
   83 & 2s2p$^5$($^1$P)3d	 & $^2$F$^o_{7/2}$   & 817.3939 & 817.0095 & 1.873$-$3 &  1.550$-$3   \\
   84 & 2s2p$^5$($^1$P)3d	 & $^2$D$^o_{5/2}$   & 817.9735 & 817.5822 & 1.074$-$3 &  9.620$-$4   \\
   85 & 2s2p$^5$($^1$P)3d	 & $^2$P$^o_{3/2}$   & 819.0236 & 818.6127 & 9.563$-$5 &  9.490$-$5   \\
   86 & 2s$^2$2p$^4$3s  	 & $^4$P$  _{1/2}$   & 829.6781 & 829.8751 & 7.003$-$8 &       $ $    \\
   87 & 2s$^2$2p$^4$($^3$P)3p	 & $^4$D$^o_{1/2}$   & 840.7645 & 840.8461 & 2.038$-$7 &  2.050$-$7   \\
   88 & 2s2p$^5$($^3$P)3s	 & $^4$P$^o_{1/2}$   & 858.3164 & 858.1844 & 1.087$-$7 &  1.030$-$7   \\
   89 & 2s2p$^5$($^3$P)3s	 & $^4$P$^o_{3/2}$   & 860.6474 & 860.5151 & 1.450$-$6 &  1.450$-$6   \\
   90 & 2s2p$^5$($^3$P)3s	 & $^2$P$^o_{1/2}$   & 862.8778 & 862.7383 & 3.711$-$7 &  3.590$-$7   \\
   91 & 2s2p$^5$($^3$P)3p	 & $^4$D$  _{1/2}$   & 868.2890 & 868.0427 & 8.322$-$8 &       $ $    \\
   92 & 2s$^2$2p$^4$($^3$P)3p	 & $^2$P$^o_{3/2}$   & 868.5910 & 868.6870 & 4.743$-$8 &  4.660$-$8   \\
   93 & 2s2p$^5$($^3$P)3p	 & $^4$D$  _{3/2}$   & 870.6954 & 870.4452 & 1.956$-$7 &       $ $    \\
   94 & 2s2p$^5$($^3$P)3p	 & $^4$P$  _{1/2}$   & 874.1049 & 873.8437 & 9.504$-$7 &       $ $    \\
   95 & 2s$^2$2p$^4$($^3$P)3d	 & $^2$D$  _{3/2}$   & 880.3596 & 880.3873 & 2.208$-$7 &       $ $    \\
   96 & 2s$^2$2p$^4$($^3$P)3d	 & $^2$D$  _{5/2}$   & 886.2175 & 886.2606 & 2.808$-$7 &       $ $    \\
   97 & 2p$^6$3s		 & $^2$S$  _{1/2}$   & 896.7062 & 896.4069 & 3.770$-$7 &       $ $    \\
   98 & 2s2p$^5$($^3$P)3p	 & $^4$P$  _{3/2}$   & 896.9621 & 896.7433 & 7.283$-$8 &       $ $    \\
   99 & 2s2p$^5$($^3$P)3p	 & $^2$D$  _{5/2}$   & 899.7352 & 899.5043 & 6.901$-$7 &       $ $    \\
  100 & 2s2p$^5$($^3$P)3p	 & $^2$D$  _{3/2}$   & 899.9941 & 899.7643 & 1.825$-$7 &       $ $    \\
  101 & 2s2p$^5$($^3$P)3p	 & $^2$S$  _{1/2}$   & 901.5742 & 901.1860 & 1.360$-$7 &       $ $    \\
  102 & 2p$^6$3p		 & $^2$P$^o_{1/2}$   & 907.0799 & 906.6040 & 3.935$-$7 &  3.290$-$7   \\
  103 & 2s2p$^5$($^3$P)3d	 & $^4$F$^o_{3/2}$   & 907.9881 & 907.6989 & 2.621$-$7 &  2.390$-$7   \\
  104 & 2s2p$^5$($^3$P)3d	 & $^4$F$^o_{5/2}$   & 910.6146 & 910.3175 & 9.255$-$7 &  8.750$-$7   \\
  105 & 2s2p$^5$($^3$P)3d	 & $^2$D$^o_{3/2}$   & 911.8530 & 911.5314 & 1.587$-$6 &  1.440$-$6   \\
  106 & 2s2p$^5$($^3$P)3d	 & $^4$D$^o_{1/2}$   & 912.6107 & 912.2002 & 9.235$-$7 &  8.030$-$7   \\
  107 & 2s2p$^5$($^3$P)3d	 & $^4$P$^o_{5/2}$   & 914.6738 & 914.3862 & 1.013$-$7 &  9.570$-$8   \\
  108 & 2s2p$^5$($^3$P)3d	 & $^2$F$^o_{7/2}$   & 917.2549 & 916.9567 & 1.914$-$6 &  1.700$-$6   \\
  109 & 2s2p$^5$($^3$P)3d	 & $^2$F$^o_{5/2}$   & 917.9829 & 917.6765 & 1.169$-$6 &  1.110$-$6   \\
  110 & 2s2p$^5$($^3$P)3d	 & $^2$P$^o_{3/2}$   & 918.1001 & 917.7930 & 2.255$-$7 &  2.190$-$7   \\
  111 & 2p$^6$3p		 & $^2$P$^o_{3/2}$   & 936.4360 & 935.8958 & 2.730$-$7 &  2.640$-$7   \\
  112 & 2p$^6$3d		 & $^2$D$  _{3/2}$   & 947.1796 & 946.5540 & 4.332$-$7 &	      \\
  113 & 2p$^6$3d		 & $^2$D$  _{5/2}$   & 954.0121 & 953.3910 & 1.309$-$6 &	      \\
\hline				  				        				   
\end{tabular}										        				   
\begin {flushleft}									        				   
\begin{tabbing} 									      
aaaaaaaaaaaaaaaaaaaaaaaaaaaaaaaaaaaa\= \kill						      		      
GRASP: Present energies with the {\sc grasp} code from 11  configurations and 113 levels \\
FAC: Present energies with the {\sc fac} code from 11  configurations and 113 levels \\
$\Omega$1: Present results from the DARC code \\
$\Omega$2: Earlier results of Goyal {\em et al.} {\cite{jam}} from the DARC code \\ % (2015)
		      		      
\end{tabbing}										      
\end {flushleft}									      
%\end{document}

\newpage			  
\begin{flushleft}			  
{\bf Table 2.} Collision strengths ($\Omega$) for transitions from the lowest three to higher excited levels of  W LXVI in the energy range 1000 to 6000~Ryd. $a{\pm}b \equiv$ $a\times$10$^{{\pm}b}$.
\newline 				  
\end{flushleft} 			  
					  
\begin{tabular}{rrllllllllll} \hline	  
    I &   J &      1000 &    1200   &    1400   &    1600   &     1800  &    2000   &	3000    &    4000   &    5000   &      6000  \\
\hline					  
    1 &   2 & 2.230$-$3 & 2.146$-$3 & 2.068$-$3 & 2.028$-$3 & 1.987$-$3 & 1.971$-$3 & 1.949$-$3 & 1.997$-$3 & 2.079$-$3 & 2.186$-$3  \\
    1 &   3 & 3.521$-$2 & 3.696$-$2 & 3.853$-$2 & 3.990$-$2 & 4.117$-$2 & 4.263$-$2 & 4.861$-$2 & 5.359$-$2 & 5.835$-$2 & 6.775$-$2  \\
    1 &   4 & 4.442$-$4 & 4.412$-$4 & 4.447$-$4 & 4.571$-$4 & 4.730$-$4 & 4.935$-$4 & 6.082$-$4 & 7.325$-$4 & 8.553$-$4 & 9.717$-$4  \\
    1 &   5 & 7.918$-$4 & 9.518$-$4 & 1.108$-$3 & 1.258$-$3 & 1.407$-$3 & 1.553$-$3 & 2.199$-$3 & 2.777$-$3 & 3.308$-$3 & 3.798$-$3  \\
    1 &   6 & 2.417$-$4 & 2.722$-$4 & 3.033$-$4 & 3.349$-$4 & 3.671$-$4 & 3.998$-$4 & 5.490$-$4 & 6.869$-$4 & 8.149$-$4 & 9.341$-$4  \\
    1 &   7 & 4.646$-$4 & 4.141$-$4 & 3.730$-$4 & 3.460$-$4 & 3.256$-$4 & 3.112$-$4 & 2.802$-$4 & 2.777$-$4 & 2.848$-$4 & 2.975$-$4  \\
    1 &   8 & 9.847$-$4 & 9.790$-$4 & 9.791$-$4 & 9.917$-$4 & 1.008$-$3 & 1.024$-$3 & 1.129$-$3 & 1.234$-$3 & 1.334$-$3 & 1.434$-$3  \\
    1 &   9 & 2.934$-$4 & 2.824$-$4 & 2.744$-$4 & 2.716$-$4 & 2.709$-$4 & 2.716$-$4 & 2.872$-$4 & 3.083$-$4 & 3.304$-$4 & 3.531$-$4  \\
    1 &  10 & 6.135$-$4 & 5.339$-$4 & 4.704$-$4 & 4.288$-$4 & 3.979$-$4 & 3.751$-$4 & 3.250$-$4 & 3.174$-$4 & 3.241$-$4 & 3.376$-$4  \\
    1 &  11 & 2.225$-$4 & 1.912$-$4 & 1.662$-$4 & 1.490$-$4 & 1.364$-$4 & 1.267$-$4 & 1.034$-$4 & 9.706$-$5 & 9.664$-$5 & 9.912$-$5  \\
    1 &  12 & 7.045$-$4 & 6.117$-$4 & 5.383$-$4 & 4.919$-$4 & 4.567$-$4 & 4.315$-$4 & 3.787$-$4 & 3.738$-$4 & 3.850$-$4 & 4.029$-$4  \\
    1 &  13 & 4.496$-$3 & 4.674$-$3 & 4.772$-$3 & 4.894$-$3 & 4.991$-$3 & 5.090$-$3 & 5.505$-$3 & 5.839$-$3 & 6.170$-$3 & 6.520$-$3  \\
    1 &  14 & 7.576$-$3 & 7.836$-$3 & 7.956$-$3 & 8.125$-$3 & 8.254$-$3 & 8.392$-$3 & 8.967$-$3 & 9.437$-$3 & 9.921$-$3 & 1.045$-$2  \\
    1 &  15 & 8.063$-$4 & 6.866$-$4 & 6.097$-$4 & 5.541$-$4 & 5.199$-$4 & 4.947$-$4 & 4.596$-$4 & 4.778$-$4 & 5.131$-$4 & 5.536$-$4  \\
    1 &  16 & 1.414$-$3 & 1.341$-$3 & 1.311$-$3 & 1.302$-$3 & 1.314$-$3 & 1.335$-$3 & 1.508$-$3 & 1.710$-$3 & 1.915$-$3 & 2.114$-$3  \\
    1 &  17 & 1.319$-$3 & 1.382$-$3 & 1.456$-$3 & 1.531$-$3 & 1.612$-$3 & 1.695$-$3 & 2.092$-$3 & 2.459$-$3 & 2.808$-$3 & 3.137$-$3  \\
    1 &  18 & 9.140$-$4 & 8.326$-$4 & 7.850$-$4 & 7.566$-$4 & 7.444$-$4 & 7.368$-$4 & 7.612$-$4 & 8.103$-$4 & 8.658$-$4 & 9.207$-$4  \\
    1 &  19 & 5.670$-$4 & 4.928$-$4 & 4.460$-$4 & 4.138$-$4 & 3.950$-$4 & 3.812$-$4 & 3.680$-$4 & 3.843$-$4 & 4.096$-$4 & 4.372$-$4  \\
    1 &  20 & 9.046$-$4 & 7.088$-$4 & 5.749$-$4 & 4.780$-$4 & 4.103$-$4 & 3.574$-$4 & 2.300$-$4 & 1.866$-$4 & 1.705$-$4 & 1.654$-$4  \\
    1 &  21 & 8.985$-$4 & 7.274$-$4 & 6.116$-$4 & 5.318$-$4 & 4.776$-$4 & 4.367$-$4 & 3.504$-$4 & 3.322$-$4 & 3.352$-$4 & 3.458$-$4  \\
    1 &  22 & 2.573$-$3 & 2.824$-$3 & 3.068$-$3 & 3.296$-$3 & 3.523$-$3 & 3.748$-$3 & 4.753$-$3 & 5.642$-$3 & 6.470$-$3 & 7.247$-$3  \\
    1 &  23 & 1.246$-$2 & 1.404$-$2 & 1.550$-$2 & 1.682$-$2 & 1.811$-$2 & 1.938$-$2 & 2.490$-$2 & 2.966$-$2 & 3.407$-$2 & 3.816$-$2  \\
    1 &  24 & 1.192$-$2 & 1.342$-$2 & 1.481$-$2 & 1.607$-$2 & 1.730$-$2 & 1.851$-$2 & 2.377$-$2 & 2.832$-$2 & 3.254$-$2 & 3.647$-$2  \\
    1 &  25 & 1.199$-$2 & 1.343$-$2 & 1.477$-$2 & 1.599$-$2 & 1.718$-$2 & 1.837$-$2 & 2.356$-$2 & 2.807$-$2 & 3.225$-$2 & 3.614$-$2  \\
    1 &  26 & 1.182$-$4 & 1.122$-$4 & 1.090$-$4 & 1.089$-$4 & 1.100$-$4 & 1.131$-$4 & 1.353$-$4 & 1.619$-$4 & 1.895$-$4 & 2.156$-$4  \\
    1 &  27 & 7.096$-$5 & 7.500$-$5 & 8.047$-$5 & 8.681$-$5 & 9.372$-$5 & 1.016$-$4 & 1.402$-$4 & 1.773$-$4 & 2.126$-$4 & 2.454$-$4  \\
    1 &  28 & 2.066$-$4 & 2.132$-$4 & 2.243$-$4 & 2.380$-$4 & 2.534$-$4 & 2.722$-$4 & 3.662$-$4 & 4.579$-$4 & 5.461$-$4 & 6.276$-$4  \\
    1 &  29 & 1.174$-$4 & 1.105$-$4 & 1.066$-$4 & 1.058$-$4 & 1.062$-$4 & 1.087$-$4 & 1.283$-$4 & 1.527$-$4 & 1.783$-$4 & 2.027$-$4  \\
    1 &  30 & 9.405$-$5 & 7.654$-$5 & 6.154$-$5 & 5.235$-$5 & 4.343$-$5 & 3.778$-$5 & 1.981$-$5 & 1.207$-$5 & 8.068$-$6 & 5.954$-$6  \\
    1 &  31 & 4.712$-$4 & 4.489$-$4 & 4.249$-$4 & 4.172$-$4 & 4.036$-$4 & 4.007$-$4 & 3.892$-$4 & 3.935$-$4 & 4.060$-$4 & 4.218$-$4  \\
    1 &  32 & 2.859$-$4 & 2.326$-$4 & 1.871$-$4 & 1.591$-$4 & 1.320$-$4 & 1.149$-$4 & 6.021$-$5 & 3.668$-$5 & 2.451$-$5 & 1.806$-$5  \\
    1 &  33 & 6.915$-$3 & 7.194$-$3 & 7.315$-$3 & 7.557$-$3 & 7.642$-$3 & 7.838$-$3 & 8.394$-$3 & 8.856$-$3 & 9.340$-$3 & 9.814$-$3  \\
    1 &  34 & 2.707$-$4 & 2.171$-$4 & 1.750$-$4 & 1.476$-$4 & 1.242$-$4 & 1.074$-$4 & 5.827$-$5 & 3.698$-$5 & 2.639$-$5 & 1.900$-$5  \\
    1 &  35 & 3.321$-$3 & 3.462$-$3 & 3.543$-$3 & 3.669$-$3 & 3.724$-$3 & 3.826$-$3 & 4.135$-$3 & 4.390$-$3 & 4.645$-$3 & 4.888$-$3  \\
    1 &  36 & 1.144$-$4 & 9.253$-$5 & 7.521$-$5 & 6.399$-$5 & 5.417$-$5 & 4.734$-$5 & 2.702$-$5 & 1.838$-$5 & 1.414$-$5 & 1.141$-$5  \\
    1 &  37 & 4.602$-$3 & 4.812$-$3 & 4.935$-$3 & 5.118$-$3 & 5.200$-$3 & 5.349$-$3 & 5.793$-$3 & 6.157$-$3 & 6.517$-$3 & 6.861$-$3  \\
    1 &  38 & 2.808$-$4 & 2.411$-$4 & 2.096$-$4 & 1.912$-$4 & 1.733$-$4 & 1.635$-$4 & 1.357$-$4 & 1.302$-$4 & 1.319$-$4 & 1.370$-$4  \\
    1 &  39 & 2.689$-$4 & 2.604$-$4 & 2.540$-$4 & 2.548$-$4 & 2.535$-$4 & 2.570$-$4 & 2.746$-$4 & 2.966$-$4 & 3.194$-$4 & 3.416$-$4  \\
    1 &  40 & 7.328$-$5 & 6.611$-$5 & 6.073$-$5 & 5.819$-$5 & 5.560$-$5 & 5.480$-$5 & 5.413$-$5 & 5.715$-$5 & 6.112$-$5 & 6.540$-$5  \\	
    1 &  41 & 4.018$-$4 & 3.775$-$4 & 3.607$-$4 & 3.567$-$4 & 3.518$-$4 & 3.544$-$4 & 3.782$-$4 & 4.133$-$4 & 4.499$-$4 & 4.857$-$4  \\
    1 &  42 & 4.529$-$4 & 5.103$-$4 & 5.679$-$4 & 6.249$-$4 & 6.808$-$4 & 7.409$-$4 & 1.011$-$3 & 1.253$-$3 & 1.480$-$3 & 1.692$-$3  \\
    1 &  43 & 2.735$-$4 & 2.514$-$4 & 2.323$-$4 & 2.231$-$4 & 2.116$-$4 & 2.073$-$4 & 1.922$-$4 & 1.910$-$4 & 1.952$-$4 & 2.018$-$4  \\
    1 &  44 & 1.584$-$3 & 1.887$-$3 & 2.178$-$3 & 2.448$-$3 & 2.709$-$3 & 2.986$-$3 & 4.187$-$3 & 5.237$-$3 & 6.209$-$3 & 7.108$-$3  \\
    1 &  45 & 2.947$-$4 & 2.635$-$4 & 2.396$-$4 & 2.274$-$4 & 2.153$-$4 & 2.105$-$4 & 2.022$-$4 & 2.105$-$4 & 2.236$-$4 & 2.383$-$4  \\
\hline	
\end{tabular}
\newpage
\begin{tabular}{rrllllllllll} \hline	  
    I &   J &      1000 &    1200   &    1400   &    1600   &     1800  &    2000   &	3000    &    4000   &    5000   &      6000  \\
\hline
    1 &  46 & 9.707$-$5 & 8.431$-$5 & 7.433$-$5 & 6.868$-$5 & 6.315$-$5 & 6.038$-$5 & 5.301$-$5 & 5.269$-$5 & 5.452$-$5 & 5.730$-$5  \\
    1 &  47 & 6.832$-$4 & 8.125$-$4 & 9.362$-$4 & 1.053$-$3 & 1.165$-$3 & 1.279$-$3 & 1.781$-$3 & 2.219$-$3 & 2.625$-$3 & 3.001$-$3  \\
    1 &  48 & 8.806$-$4 & 1.028$-$3 & 1.170$-$3 & 1.305$-$3 & 1.435$-$3 & 1.570$-$3 & 2.167$-$3 & 2.690$-$3 & 3.176$-$3 & 3.627$-$3  \\
    1 &  49 & 8.850$-$5 & 6.725$-$5 & 5.267$-$5 & 4.247$-$5 & 3.474$-$5 & 2.904$-$5 & 1.443$-$5 & 9.080$-$6 & 6.727$-$6 & 5.568$-$6  \\
    1 &  50 & 3.003$-$4 & 2.745$-$4 & 2.606$-$4 & 2.544$-$4 & 2.521$-$4 & 2.539$-$4 & 2.814$-$4 & 3.192$-$4 & 3.596$-$4 & 3.989$-$4  \\
    1 &  51 & 2.237$-$3 & 2.469$-$3 & 2.693$-$3 & 2.910$-$3 & 3.115$-$3 & 3.333$-$3 & 4.288$-$3 & 5.123$-$3 & 5.908$-$3 & 6.633$-$3  \\
    1 &  52 & 3.611$-$4 & 2.736$-$4 & 2.134$-$4 & 1.712$-$4 & 1.392$-$4 & 1.154$-$4 & 5.385$-$5 & 3.050$-$5 & 1.959$-$5 & 1.368$-$5  \\
    1 &  53 & 2.704$-$3 & 3.070$-$3 & 3.406$-$3 & 3.725$-$3 & 4.021$-$3 & 4.315$-$3 & 5.598$-$3 & 6.700$-$3 & 7.729$-$3 & 8.678$-$3  \\
    1 &  54 & 6.845$-$3 & 7.780$-$3 & 8.642$-$3 & 9.448$-$3 & 1.020$-$2 & 1.098$-$2 & 1.430$-$2 & 1.716$-$2 & 1.983$-$2 & 2.228$-$2  \\
    1 &  55 & 5.511$-$3 & 6.271$-$3 & 6.969$-$3 & 7.625$-$3 & 8.239$-$3 & 8.855$-$3 & 1.153$-$2 & 1.382$-$2 & 1.595$-$2 & 1.792$-$2  \\
    1 &  56 & 5.147$-$4 & 4.030$-$4 & 3.280$-$4 & 2.751$-$4 & 2.360$-$4 & 2.064$-$4 & 1.338$-$4 & 1.092$-$4 & 1.003$-$4 & 9.747$-$5  \\
    1 &  57 & 2.992$-$4 & 2.590$-$4 & 2.346$-$4 & 2.195$-$4 & 2.094$-$4 & 2.034$-$4 & 1.976$-$4 & 2.056$-$4 & 2.175$-$4 & 2.303$-$4  \\
    1 &  58 & 2.489$-$4 & 1.985$-$4 & 1.651$-$4 & 1.418$-$4 & 1.248$-$4 & 1.120$-$4 & 8.213$-$5 & 7.341$-$5 & 7.156$-$5 & 7.227$-$5  \\
    1 &  59 & 4.392$-$4 & 4.020$-$4 & 3.831$-$4 & 3.748$-$4 & 3.710$-$4 & 3.719$-$4 & 3.948$-$4 & 4.259$-$4 & 4.583$-$4 & 4.895$-$4  \\
    1 &  60 & 3.800$-$4 & 2.974$-$4 & 2.421$-$4 & 2.032$-$4 & 1.744$-$4 & 1.527$-$4 & 9.961$-$5 & 8.185$-$5 & 7.564$-$5 & 7.388$-$5  \\
    1 &  61 & 3.867$-$4 & 3.363$-$4 & 3.062$-$4 & 2.880$-$4 & 2.758$-$4 & 2.690$-$4 & 2.643$-$4 & 2.765$-$4 & 2.933$-$4 & 3.110$-$4  \\
    1 &  62 & 2.713$-$4 & 2.488$-$4 & 2.376$-$4 & 2.330$-$4 & 2.317$-$4 & 2.337$-$4 & 2.583$-$4 & 2.901$-$4 & 3.234$-$4 & 3.555$-$4  \\
    1 &  63 & 1.962$-$4 & 1.712$-$4 & 1.561$-$4 & 1.470$-$4 & 1.415$-$4 & 1.387$-$4 & 1.432$-$4 & 1.582$-$4 & 1.765$-$4 & 1.949$-$4  \\
    1 &  64 & 1.819$-$4 & 1.487$-$4 & 1.220$-$4 & 1.031$-$4 & 8.640$-$5 & 7.514$-$5 & 3.968$-$5 & 2.427$-$5 & 1.630$-$5 & 1.167$-$5  \\
    1 &  65 & 3.342$-$4 & 4.212$-$4 & 5.099$-$4 & 5.946$-$4 & 6.793$-$4 & 7.674$-$4 & 1.167$-$3 & 1.522$-$3 & 1.853$-$3 & 2.160$-$3  \\
    1 &  66 & 3.581$-$4 & 4.313$-$4 & 5.079$-$4 & 5.824$-$4 & 6.564$-$4 & 7.385$-$4 & 1.108$-$3 & 1.441$-$3 & 1.753$-$3 & 2.042$-$3  \\
    1 &  67 & 2.121$-$4 & 2.753$-$4 & 3.390$-$4 & 3.999$-$4 & 4.603$-$4 & 5.218$-$4 & 8.006$-$4 & 1.048$-$3 & 1.277$-$3 & 1.489$-$3  \\
    1 &  68 & 4.029$-$4 & 4.966$-$4 & 5.936$-$4 & 6.864$-$4 & 7.790$-$4 & 8.798$-$4 & 1.333$-$3 & 1.738$-$3 & 2.116$-$3 & 2.466$-$3  \\
    1 &  69 & 1.871$-$4 & 2.151$-$4 & 2.459$-$4 & 2.773$-$4 & 3.093$-$4 & 3.442$-$4 & 5.073$-$4 & 6.570$-$4 & 7.980$-$4 & 9.295$-$4  \\
    1 &  70 & 5.156$-$5 & 4.472$-$5 & 3.958$-$5 & 3.635$-$5 & 3.368$-$5 & 3.239$-$5 & 3.047$-$5 & 3.275$-$5 & 3.644$-$5 & 4.053$-$5  \\
    1 &  71 & 1.015$-$4 & 8.352$-$5 & 7.063$-$5 & 6.151$-$5 & 5.464$-$5 & 4.959$-$5 & 3.681$-$5 & 3.289$-$5 & 3.189$-$5 & 3.195$-$5  \\
    1 &  72 & 2.512$-$4 & 2.253$-$4 & 2.085$-$4 & 1.978$-$4 & 1.908$-$4 & 1.869$-$4 & 1.846$-$4 & 1.938$-$4 & 2.060$-$4 & 2.184$-$4  \\
    1 &  73 & 1.240$-$3 & 1.340$-$3 & 1.439$-$3 & 1.530$-$3 & 1.617$-$3 & 1.700$-$3 & 2.043$-$3 & 2.317$-$3 & 2.556$-$3 & 2.764$-$3  \\
    1 &  74 & 5.688$-$4 & 5.698$-$4 & 5.793$-$4 & 5.927$-$4 & 6.087$-$4 & 6.267$-$4 & 7.142$-$4 & 7.952$-$4 & 8.695$-$4 & 9.361$-$4  \\
    1 &  75 & 4.093$-$4 & 4.551$-$4 & 4.978$-$4 & 5.359$-$4 & 5.714$-$4 & 6.047$-$4 & 7.380$-$4 & 8.418$-$4 & 9.307$-$4 & 1.008$-$3  \\
    1 &  76 & 3.703$-$4 & 3.313$-$4 & 3.060$-$4 & 2.897$-$4 & 2.792$-$4 & 2.731$-$4 & 2.690$-$4 & 2.821$-$4 & 2.998$-$4 & 3.176$-$4  \\
    1 &  77 & 2.625$-$4 & 2.395$-$4 & 2.254$-$4 & 2.168$-$4 & 2.119$-$4 & 2.097$-$4 & 2.141$-$4 & 2.282$-$4 & 2.445$-$4 & 2.602$-$4  \\
    1 &  78 & 4.402$-$4 & 3.427$-$4 & 2.742$-$4 & 2.236$-$4 & 1.860$-$4 & 1.567$-$4 & 7.841$-$5 & 4.685$-$5 & 3.114$-$5 & 2.182$-$5  \\
    1 &  79 & 1.044$-$3 & 1.110$-$3 & 1.179$-$3 & 1.245$-$3 & 1.310$-$3 & 1.372$-$3 & 1.637$-$3 & 1.854$-$3 & 2.044$-$3 & 2.210$-$3  \\
    1 &  80 & 1.251$-$3 & 1.397$-$3 & 1.534$-$3 & 1.655$-$3 & 1.768$-$3 & 1.873$-$3 & 2.294$-$3 & 2.621$-$3 & 2.900$-$3 & 3.142$-$3  \\
    1 &  81 & 1.348$-$3 & 1.555$-$3 & 1.740$-$3 & 1.900$-$3 & 2.048$-$3 & 2.183$-$3 & 2.711$-$3 & 3.111$-$3 & 3.449$-$3 & 3.741$-$3  \\
    1 &  82 & 4.228$-$4 & 4.743$-$4 & 5.221$-$4 & 5.642$-$4 & 6.037$-$4 & 6.402$-$4 & 7.859$-$4 & 8.984$-$4 & 9.941$-$4 & 1.077$-$3  \\
    1 &  83 & 1.447$-$3 & 1.599$-$3 & 1.743$-$3 & 1.873$-$3 & 1.995$-$3 & 2.109$-$3 & 2.571$-$3 & 2.934$-$3 & 3.244$-$3 & 3.514$-$3  \\
    1 &  84 & 8.712$-$4 & 9.404$-$4 & 1.010$-$3 & 1.074$-$3 & 1.136$-$3 & 1.194$-$3 & 1.439$-$3 & 1.635$-$3 & 1.805$-$3 & 1.954$-$3  \\
    1 &  85 & 1.806$-$4 & 1.423$-$4 & 1.154$-$4 & 9.563$-$5 & 8.097$-$5 & 6.961$-$5 & 3.972$-$5 & 2.814$-$5 & 2.274$-$5 & 1.977$-$5  \\
    1 &  86 & 9.949$-$8 & 8.545$-$8 & 7.645$-$8 & 7.003$-$8 & 6.525$-$8 & 6.211$-$8 & 5.602$-$8 & 5.696$-$8 & 6.063$-$8 & 6.464$-$8  \\
    1 &  87 & 3.629$-$7 & 2.906$-$7 & 2.383$-$7 & 2.038$-$7 & 1.735$-$7 & 1.559$-$7 & 1.026$-$7 & 8.385$-$8 & 7.659$-$8 & 7.223$-$8  \\
    1 &  88 & 1.406$-$7 & 1.256$-$7 & 1.151$-$7 & 1.087$-$7 & 1.034$-$7 & 1.011$-$7 & 9.682$-$8 & 9.990$-$8 & 1.051$-$7 & 1.108$-$7  \\
    1 &  89 & 1.390$-$6 & 1.409$-$6 & 1.424$-$6 & 1.450$-$6 & 1.459$-$6 & 1.493$-$6 & 1.589$-$6 & 1.682$-$6 & 1.775$-$6 & 1.862$-$6  \\
    1 &  90 & 4.935$-$7 & 4.355$-$7 & 3.948$-$7 & 3.711$-$7 & 3.495$-$7 & 3.413$-$7 & 3.199$-$7 & 3.243$-$7 & 3.368$-$7 & 3.504$-$7  \\
\hline	
\end{tabular}
\newpage
\begin{tabular}{rrllllllllll} \hline	  
    I &   J &      1000 &    1200   &    1400   &    1600   &     1800  &    2000   &	3000    &    4000   &    5000   &      6000  \\
\hline
    1 &  91 & 1.082$-$7 & 9.652$-$8 & 8.820$-$8 & 8.322$-$8 & 7.888$-$8 & 7.701$-$8 & 7.437$-$8 & 7.836$-$8 & 8.520$-$8 & 9.191$-$8  \\
    1 &  92 & 8.382$-$8 & 6.702$-$8 & 5.515$-$8 & 4.743$-$8 & 4.136$-$8 & 3.750$-$8 & 2.754$-$8 & 2.473$-$8 & 2.422$-$8 & 2.456$-$8  \\
    1 &  93 & 2.389$-$7 & 2.157$-$7 & 2.018$-$7 & 1.956$-$7 & 1.923$-$7 & 1.935$-$7 & 2.164$-$7 & 2.501$-$7 & 2.871$-$7 & 3.233$-$7  \\
    1 &  94 & 1.156$-$6 & 1.067$-$6 & 9.924$-$7 & 9.504$-$7 & 9.019$-$7 & 8.830$-$7 & 8.175$-$7 & 8.159$-$7 & 8.548$-$7 & 8.865$-$7  \\
    1 &  95 & 3.860$-$7 & 3.094$-$7 & 2.583$-$7 & 2.208$-$7 & 1.922$-$7 & 1.721$-$7 & 1.159$-$7 & 9.358$-$8 & 8.279$-$8 & 7.627$-$8  \\
    1 &  96 & 2.967$-$7 & 2.800$-$7 & 2.771$-$7 & 2.808$-$7 & 2.881$-$7 & 2.995$-$7 & 3.665$-$7 & 4.358$-$7 & 5.039$-$7 & 5.683$-$7  \\
    1 &  97 & 3.647$-$7 & 3.661$-$7 & 3.683$-$7 & 3.770$-$7 & 3.824$-$7 & 3.940$-$7 & 4.484$-$7 & 5.079$-$7 & 5.731$-$7 & 6.327$-$7  \\
    1 &  98 & 9.524$-$8 & 8.435$-$8 & 7.729$-$8 & 7.283$-$8 & 6.983$-$8 & 6.842$-$8 & 6.835$-$8 & 7.316$-$8 & 7.978$-$8 & 8.647$-$8  \\
    1 &  99 & 4.754$-$7 & 5.391$-$7 & 6.117$-$7 & 6.901$-$7 & 7.696$-$7 & 8.536$-$7 & 1.260$-$6 & 1.636$-$6 & 1.990$-$6 & 2.325$-$6  \\
    1 & 100 & 2.156$-$7 & 1.983$-$7 & 1.875$-$7 & 1.825$-$7 & 1.813$-$7 & 1.827$-$7 & 2.086$-$7 & 2.446$-$7 & 2.834$-$7 & 3.217$-$7  \\
    1 & 101 & 1.440$-$7 & 1.393$-$7 & 1.362$-$7 & 1.360$-$7 & 1.357$-$7 & 1.381$-$7 & 1.543$-$7 & 1.763$-$7 & 2.013$-$7 & 2.254$-$7  \\
    1 & 102 & 3.868$-$7 & 3.904$-$7 & 3.930$-$7 & 3.935$-$7 & 3.944$-$7 & 3.959$-$7 & 3.932$-$7 & 3.884$-$7 & 3.848$-$7 & 3.824$-$7  \\
    1 & 103 & 3.022$-$7 & 2.793$-$7 & 2.684$-$7 & 2.621$-$7 & 2.603$-$7 & 2.603$-$7 & 2.694$-$7 & 2.814$-$7 & 2.931$-$7 & 3.042$-$7  \\
    1 & 104 & 9.708$-$7 & 9.336$-$7 & 9.224$-$7 & 9.255$-$7 & 9.388$-$7 & 9.572$-$7 & 1.076$-$6 & 1.194$-$6 & 1.303$-$6 & 1.404$-$6  \\
    1 & 105 & 1.601$-$6 & 1.583$-$6 & 1.582$-$6 & 1.587$-$6 & 1.595$-$6 & 1.611$-$6 & 1.687$-$6 & 1.763$-$6 & 1.842$-$6 & 1.914$-$6  \\
    1 & 106 & 9.558$-$7 & 9.421$-$7 & 9.330$-$7 & 9.235$-$7 & 9.158$-$7 & 9.115$-$7 & 8.907$-$7 & 8.824$-$7 & 8.857$-$7 & 8.915$-$7  \\
    1 & 107 & 1.440$-$7 & 1.214$-$7 & 1.087$-$7 & 1.013$-$7 & 9.726$-$8 & 9.519$-$8 & 9.566$-$8 & 1.013$-$7 & 1.079$-$7 & 1.145$-$7  \\
    1 & 108 & 1.624$-$6 & 1.712$-$6 & 1.814$-$6 & 1.914$-$6 & 2.018$-$6 & 2.115$-$6 & 2.556$-$6 & 2.912$-$6 & 3.217$-$6 & 3.487$-$6  \\
    1 & 109 & 1.128$-$6 & 1.123$-$6 & 1.141$-$6 & 1.169$-$6 & 1.205$-$6 & 1.243$-$6 & 1.443$-$6 & 1.622$-$6 & 1.781$-$6 & 1.924$-$6  \\
    1 & 110 & 3.628$-$7 & 2.995$-$7 & 2.557$-$7 & 2.255$-$7 & 2.039$-$7 & 1.892$-$7 & 1.554$-$7 & 1.487$-$7 & 1.500$-$7 & 1.543$-$7  \\
    1 & 111 & 3.559$-$7 & 3.200$-$7 & 2.912$-$7 & 2.730$-$7 & 2.578$-$7 & 2.499$-$7 & 2.279$-$7 & 2.246$-$7 & 2.268$-$7 & 2.317$-$7  \\
    1 & 112 & 5.862$-$7 & 5.102$-$7 & 4.658$-$7 & 4.332$-$7 & 4.145$-$7 & 4.013$-$7 & 3.788$-$7 & 3.841$-$7 & 3.999$-$7 & 4.190$-$7  \\
    1 & 113 & 1.248$-$6 & 1.243$-$6 & 1.272$-$6 & 1.309$-$6 & 1.358$-$6 & 1.412$-$6 & 1.682$-$6 & 1.940$-$6 & 2.191$-$6 & 2.430$-$6  \\
    2 &   3 & 2.281$-$2 & 2.396$-$2 & 2.505$-$2 & 2.605$-$2 & 2.700$-$2 & 2.798$-$2 & 3.260$-$2 & 3.703$-$2 & 4.162$-$2 & 5.064$-$2  \\
    2 &   4 & 3.589$-$8 & 2.891$-$8 & 2.379$-$8 & 1.998$-$8 & 1.704$-$8 & 1.471$-$8 & 8.195$-$9 & 5.409$-$9 & 3.995$-$9 & 3.101$-$9  \\
    2 &   5 & 5.392$-$8 & 5.613$-$8 & 5.958$-$8 & 6.330$-$8 & 6.770$-$8 & 7.141$-$8 & 9.076$-$8 & 1.088$-$7 & 1.254$-$7 & 1.409$-$7  \\
    2 &   6 & 8.630$-$8 & 9.099$-$8 & 9.650$-$8 & 1.027$-$7 & 1.095$-$7 & 1.152$-$7 & 1.458$-$7 & 1.759$-$7 & 2.043$-$7 & 2.316$-$7  \\
    2 &   7 & 7.631$-$8 & 6.880$-$8 & 6.360$-$8 & 6.037$-$8 & 5.811$-$8 & 5.651$-$8 & 5.465$-$8 & 5.616$-$8 & 5.869$-$8 & 6.199$-$8  \\	
    2 &   8 & 8.735$-$8 & 8.277$-$8 & 8.030$-$8 & 7.915$-$8 & 7.897$-$8 & 7.885$-$8 & 8.324$-$8 & 8.921$-$8 & 9.536$-$8 & 1.020$-$7  \\
    2 &   9 & 1.976$-$7 & 1.892$-$7 & 1.832$-$7 & 1.790$-$7 & 1.771$-$7 & 1.745$-$7 & 1.746$-$7 & 1.782$-$7 & 1.853$-$7 & 1.926$-$7  \\
    2 &  10 & 8.682$-$8 & 7.715$-$8 & 7.085$-$8 & 6.714$-$8 & 6.487$-$8 & 6.325$-$8 & 6.287$-$8 & 6.621$-$8 & 7.042$-$8 & 7.527$-$8  \\
    2 &  11 & 5.110$-$7 & 5.207$-$7 & 5.267$-$7 & 5.343$-$7 & 5.426$-$7 & 5.482$-$7 & 5.828$-$7 & 6.120$-$7 & 6.442$-$7 & 6.748$-$7  \\
    2 &  12 & 6.056$-$8 & 4.724$-$8 & 3.763$-$8 & 3.073$-$8 & 2.561$-$8 & 2.151$-$8 & 1.094$-$8 & 6.570$-$9 & 4.448$-$9 & 3.239$-$9  \\
    2 &  13 & 3.562$-$7 & 3.207$-$7 & 2.940$-$7 & 2.797$-$7 & 2.687$-$7 & 2.615$-$7 & 2.524$-$7 & 2.585$-$7 & 2.700$-$7 & 2.830$-$7  \\
    2 &  14 & 5.616$-$7 & 4.710$-$7 & 4.026$-$7 & 3.586$-$7 & 3.250$-$7 & 2.993$-$7 & 2.377$-$7 & 2.160$-$7 & 2.096$-$7 & 2.082$-$7  \\
    2 &  15 & 1.095$-$6 & 1.178$-$6 & 1.263$-$6 & 1.342$-$6 & 1.424$-$6 & 1.497$-$6 & 1.839$-$6 & 2.149$-$6 & 2.437$-$6 & 2.709$-$6  \\
    2 &  16 & 1.422$-$7 & 1.145$-$7 & 9.533$-$8 & 8.143$-$8 & 7.190$-$8 & 6.428$-$8 & 4.668$-$8 & 4.128$-$8 & 3.979$-$8 & 3.968$-$8  \\
    2 &  17 & 3.561$-$7 & 3.739$-$7 & 3.937$-$7 & 4.134$-$7 & 4.354$-$7 & 4.546$-$7 & 5.524$-$7 & 6.442$-$7 & 7.310$-$7 & 8.133$-$7  \\
    2 &  18 & 7.339$-$8 & 5.865$-$8 & 4.920$-$8 & 4.310$-$8 & 3.912$-$8 & 3.629$-$8 & 3.106$-$8 & 3.068$-$8 & 3.157$-$8 & 3.286$-$8  \\
    2 &  19 & 6.462$-$7 & 6.846$-$7 & 7.284$-$7 & 7.720$-$7 & 8.192$-$7 & 8.617$-$7 & 1.067$-$6 & 1.255$-$6 & 1.430$-$6 & 1.594$-$6  \\
    2 &  20 & 1.351$-$7 & 1.105$-$7 & 9.420$-$8 & 8.298$-$8 & 7.560$-$8 & 7.002$-$8 & 5.921$-$8 & 5.799$-$8 & 5.951$-$8 & 6.195$-$8  \\
    2 &  21 & 8.789$-$8 & 6.438$-$8 & 4.878$-$8 & 3.793$-$8 & 3.050$-$8 & 2.482$-$8 & 1.147$-$8 & 6.743$-$9 & 4.597$-$9 & 3.443$-$9  \\
    2 &  22 & 2.038$-$7 & 1.775$-$7 & 1.602$-$7 & 1.476$-$7 & 1.399$-$7 & 1.330$-$7 & 1.202$-$7 & 1.185$-$7 & 1.209$-$7 & 1.246$-$7  \\
\hline	
\end{tabular}
\newpage
\begin{tabular}{rrllllllllll} \hline	  
    I &   J &      1000 &    1200   &    1400   &    1600   &     1800  &    2000   &	3000    &    4000   &    5000   &      6000  \\
\hline
    2 &  23 & 2.929$-$7 & 2.453$-$7 & 2.118$-$7 & 1.885$-$7 & 1.721$-$7 & 1.587$-$7 & 1.256$-$7 & 1.121$-$7 & 1.060$-$7 & 1.031$-$7  \\
    2 &  24 & 4.742$-$7 & 4.331$-$7 & 4.106$-$7 & 3.969$-$7 & 3.928$-$7 & 3.895$-$7 & 4.077$-$7 & 4.402$-$7 & 4.768$-$7 & 5.132$-$7  \\
    2 &  25 & 3.662$-$7 & 2.952$-$7 & 2.453$-$7 & 2.096$-$7 & 1.844$-$7 & 1.636$-$7 & 1.108$-$7 & 8.777$-$8 & 7.581$-$8 & 6.844$-$8  \\
    2 &  26 & 1.136$-$4 & 9.450$-$5 & 8.068$-$5 & 7.073$-$5 & 6.296$-$5 & 5.755$-$5 & 4.403$-$5 & 4.149$-$5 & 4.275$-$5 & 4.514$-$5  \\
    2 &  27 & 2.240$-$4 & 2.818$-$4 & 3.381$-$4 & 3.915$-$4 & 4.439$-$4 & 4.957$-$4 & 7.224$-$4 & 9.220$-$4 & 1.103$-$3 & 1.271$-$3  \\
    2 &  28 & 1.635$-$4 & 1.316$-$4 & 1.078$-$4 & 9.008$-$5 & 7.572$-$5 & 6.502$-$5 & 3.371$-$5 & 2.051$-$5 & 1.394$-$5 & 9.769$-$6  \\
    2 &  29 & 4.394$-$4 & 5.527$-$4 & 6.628$-$4 & 7.666$-$4 & 8.696$-$4 & 9.717$-$4 & 1.416$-$3 & 1.808$-$3 & 2.165$-$3 & 2.493$-$3  \\
    2 &  30 & 1.103$-$4 & 8.829$-$5 & 7.096$-$5 & 5.976$-$5 & 4.976$-$5 & 4.277$-$5 & 2.249$-$5 & 1.370$-$5 & 9.209$-$6 & 6.757$-$6  \\
    2 &  31 & 2.950$-$4 & 2.752$-$4 & 2.619$-$4 & 2.568$-$4 & 2.525$-$4 & 2.512$-$4 & 2.596$-$4 & 2.768$-$4 & 2.958$-$4 & 3.156$-$4  \\
    2 &  32 & 5.248$-$4 & 5.227$-$4 & 5.271$-$4 & 5.397$-$4 & 5.510$-$4 & 5.634$-$4 & 6.316$-$4 & 6.970$-$4 & 7.571$-$4 & 8.152$-$4  \\
    2 &  33 & 2.804$-$4 & 2.535$-$4 & 2.341$-$4 & 2.243$-$4 & 2.156$-$4 & 2.108$-$4 & 2.063$-$4 & 2.146$-$4 & 2.264$-$4 & 2.400$-$4  \\
    2 &  34 & 9.685$-$6 & 7.990$-$6 & 6.773$-$6 & 6.001$-$6 & 5.353$-$6 & 4.913$-$6 & 3.811$-$6 & 3.512$-$6 & 3.482$-$6 & 3.551$-$6  \\
    2 &  35 & 6.352$-$6 & 6.300$-$6 & 6.347$-$6 & 6.511$-$6 & 6.668$-$6 & 6.847$-$6 & 7.805$-$6 & 8.712$-$6 & 9.534$-$6 & 1.031$-$5  \\
    2 &  36 & 4.286$-$5 & 3.904$-$5 & 3.595$-$5 & 3.437$-$5 & 3.264$-$5 & 3.176$-$5 & 2.928$-$5 & 2.888$-$5 & 2.947$-$5 & 3.017$-$5  \\
    2 &  37 & 2.252$-$5 & 2.172$-$5 & 2.140$-$5 & 2.163$-$5 & 2.187$-$5 & 2.229$-$5 & 2.491$-$5 & 2.765$-$5 & 3.024$-$5 & 3.269$-$5  \\
    2 &  38 & 2.738$-$4 & 2.190$-$4 & 1.795$-$4 & 1.536$-$4 & 1.320$-$4 & 1.169$-$4 & 7.651$-$5 & 6.218$-$5 & 5.683$-$5 & 5.487$-$5  \\
    2 &  39 & 2.358$-$4 & 2.066$-$4 & 1.853$-$4 & 1.739$-$4 & 1.636$-$4 & 1.580$-$4 & 1.479$-$4 & 1.515$-$4 & 1.590$-$4 & 1.680$-$4  \\
    2 &  40 & 4.431$-$4 & 4.347$-$4 & 4.266$-$4 & 4.277$-$4 & 4.239$-$4 & 4.260$-$4 & 4.353$-$4 & 4.509$-$4 & 4.723$-$4 & 4.908$-$4  \\
    2 &  41 & 3.682$-$4 & 2.870$-$4 & 2.282$-$4 & 1.887$-$4 & 1.555$-$4 & 1.320$-$4 & 6.612$-$5 & 3.918$-$5 & 2.609$-$5 & 1.834$-$5  \\
    2 &  42 & 1.383$-$6 & 1.320$-$6 & 1.306$-$6 & 1.318$-$6 & 1.353$-$6 & 1.403$-$6 & 1.701$-$6 & 2.022$-$6 & 2.338$-$6 & 2.636$-$6  \\
    2 &  43 & 2.234$-$4 & 1.990$-$4 & 1.818$-$4 & 1.731$-$4 & 1.654$-$4 & 1.617$-$4 & 1.580$-$4 & 1.652$-$4 & 1.753$-$4 & 1.864$-$4  \\
    2 &  44 & 3.357$-$6 & 3.066$-$6 & 2.905$-$6 & 2.838$-$6 & 2.804$-$6 & 2.805$-$6 & 2.966$-$6 & 3.201$-$6 & 3.441$-$6 & 3.672$-$6  \\
    2 &  45 & 3.336$-$4 & 3.270$-$4 & 3.269$-$4 & 3.338$-$4 & 3.402$-$4 & 3.486$-$4 & 3.955$-$4 & 4.415$-$4 & 4.838$-$4 & 5.238$-$4  \\
    2 &  46 & 7.338$-$3 & 7.623$-$3 & 7.808$-$3 & 8.063$-$3 & 8.186$-$3 & 8.376$-$3 & 8.993$-$3 & 9.511$-$3 & 1.007$-$2 & 1.052$-$2  \\
    2 &  47 & 9.839$-$6 & 8.898$-$6 & 8.340$-$6 & 8.029$-$6 & 7.894$-$6 & 7.852$-$6 & 8.391$-$6 & 9.346$-$6 & 1.040$-$5 & 1.144$-$5  \\	
    2 &  48 & 4.375$-$6 & 3.650$-$6 & 3.222$-$6 & 2.986$-$6 & 2.873$-$6 & 2.845$-$6 & 3.177$-$6 & 3.771$-$6 & 4.411$-$6 & 5.034$-$6  \\
    2 &  49 & 4.089$-$4 & 3.579$-$4 & 3.257$-$4 & 3.057$-$4 & 2.941$-$4 & 2.875$-$4 & 2.930$-$4 & 3.214$-$4 & 3.561$-$4 & 3.916$-$4  \\
    2 &  50 & 6.272$-$4 & 5.790$-$4 & 5.544$-$4 & 5.436$-$4 & 5.440$-$4 & 5.501$-$4 & 6.185$-$4 & 7.063$-$4 & 7.976$-$4 & 8.852$-$4  \\
    2 &  51 & 3.479$-$4 & 2.986$-$4 & 2.680$-$4 & 2.501$-$4 & 2.382$-$4 & 2.312$-$4 & 2.253$-$4 & 2.358$-$4 & 2.501$-$4 & 2.649$-$4  \\
    2 &  52 & 5.215$-$4 & 4.841$-$4 & 4.652$-$4 & 4.593$-$4 & 4.574$-$4 & 4.605$-$4 & 4.939$-$4 & 5.352$-$4 & 5.763$-$4 & 6.152$-$4  \\
    2 &  53 & 3.566$-$4 & 3.079$-$4 & 2.766$-$4 & 2.568$-$4 & 2.446$-$4 & 2.371$-$4 & 2.354$-$4 & 2.553$-$4 & 2.813$-$4 & 3.084$-$4  \\
    2 &  54 & 3.372$-$4 & 2.800$-$4 & 2.433$-$4 & 2.205$-$4 & 2.048$-$4 & 1.946$-$4 & 1.778$-$4 & 1.812$-$4 & 1.898$-$4 & 1.998$-$4  \\
    2 &  55 & 3.794$-$4 & 3.141$-$4 & 2.716$-$4 & 2.438$-$4 & 2.259$-$4 & 2.139$-$4 & 1.988$-$4 & 2.106$-$4 & 2.301$-$4 & 2.513$-$4  \\
    2 &  56 & 4.385$-$4 & 3.283$-$4 & 2.545$-$4 & 2.038$-$4 & 1.668$-$4 & 1.395$-$4 & 7.226$-$5 & 4.868$-$5 & 3.856$-$5 & 3.354$-$5  \\
    2 &  57 & 3.531$-$4 & 2.883$-$4 & 2.469$-$4 & 2.202$-$4 & 2.020$-$4 & 1.893$-$4 & 1.657$-$4 & 1.654$-$4 & 1.713$-$4 & 1.792$-$4  \\
    2 &  58 & 1.306$-$3 & 1.380$-$3 & 1.467$-$3 & 1.554$-$3 & 1.649$-$3 & 1.745$-$3 & 2.196$-$3 & 2.607$-$3 & 2.996$-$3 & 3.355$-$3  \\
    2 &  59 & 5.574$-$4 & 4.136$-$4 & 3.172$-$4 & 2.505$-$4 & 2.017$-$4 & 1.654$-$4 & 7.506$-$5 & 4.207$-$5 & 2.691$-$5 & 1.860$-$5  \\
    2 &  60 & 3.765$-$4 & 3.027$-$4 & 2.549$-$4 & 2.232$-$4 & 2.013$-$4 & 1.856$-$4 & 1.532$-$4 & 1.484$-$4 & 1.513$-$4 & 1.569$-$4  \\
    2 &  61 & 3.411$-$4 & 3.108$-$4 & 2.966$-$4 & 2.922$-$4 & 2.915$-$4 & 2.943$-$4 & 3.206$-$4 & 3.506$-$4 & 3.794$-$4 & 4.062$-$4  \\
    2 &  62 & 1.486$-$2 & 1.705$-$2 & 1.907$-$2 & 2.087$-$2 & 2.265$-$2 & 2.436$-$2 & 3.174$-$2 & 3.809$-$2 & 4.395$-$2 & 4.932$-$2  \\
    2 &  63 & 8.151$-$3 & 9.355$-$3 & 1.047$-$2 & 1.148$-$2 & 1.244$-$2 & 1.338$-$2 & 1.743$-$2 & 2.092$-$2 & 2.413$-$2 & 2.708$-$2  \\
    2 &  64 & 2.292$-$6 & 1.712$-$6 & 1.325$-$6 & 1.056$-$6 & 8.608$-$7 & 7.154$-$7 & 3.579$-$7 & 2.314$-$7 & 1.761$-$7 & 1.476$-$7  \\
    2 &  65 & 1.106$-$5 & 1.238$-$5 & 1.364$-$5 & 1.478$-$5 & 1.593$-$5 & 1.704$-$5 & 2.189$-$5 & 2.611$-$5 & 3.003$-$5 & 3.362$-$5  \\
    2 &  66 & 1.114$-$6 & 1.065$-$6 & 1.052$-$6 & 1.060$-$6 & 1.073$-$6 & 1.094$-$6 & 1.209$-$6 & 1.322$-$6 & 1.429$-$6 & 1.527$-$6  \\
    2 &  67 & 2.226$-$6 & 2.476$-$6 & 2.730$-$6 & 2.966$-$6 & 3.199$-$6 & 3.436$-$6 & 4.485$-$6 & 5.421$-$6 & 6.293$-$6 & 7.096$-$6  \\
\hline	
\end{tabular}
\newpage
\begin{tabular}{rrllllllllll} \hline	  
    I &   J &      1000 &    1200   &    1400   &    1600   &     1800  &    2000   &	3000    &    4000   &    5000   &      6000  \\
\hline
    2 &  68 & 9.261$-$7 & 7.078$-$7 & 5.622$-$7 & 4.613$-$7 & 3.859$-$7 & 3.330$-$7 & 1.990$-$7 & 1.538$-$7 & 1.365$-$7 & 1.294$-$7  \\
    2 &  69 & 7.107$-$6 & 8.277$-$6 & 9.447$-$6 & 1.053$-$5 & 1.163$-$5 & 1.276$-$5 & 1.771$-$5 & 2.208$-$5 & 2.613$-$5 & 2.985$-$5  \\
    2 &  70 & 1.898$-$5 & 2.196$-$5 & 2.475$-$5 & 2.733$-$5 & 2.976$-$5 & 3.221$-$5 & 4.267$-$5 & 5.174$-$5 & 6.011$-$5 & 6.778$-$5  \\
    2 &  71 & 1.339$-$7 & 1.144$-$7 & 1.013$-$7 & 9.383$-$8 & 8.701$-$8 & 8.363$-$8 & 7.447$-$8 & 7.279$-$8 & 7.397$-$8 & 7.557$-$8  \\
    2 &  72 & 1.169$-$7 & 9.426$-$8 & 7.967$-$8 & 6.983$-$8 & 6.242$-$8 & 5.748$-$8 & 4.451$-$8 & 3.984$-$8 & 3.795$-$8 & 3.695$-$8  \\
    2 &  73 & 8.133$-$8 & 6.208$-$8 & 5.020$-$8 & 4.219$-$8 & 3.648$-$8 & 3.270$-$8 & 2.279$-$8 & 1.919$-$8 & 1.751$-$8 & 1.658$-$8  \\
    2 &  74 & 1.611$-$7 & 1.492$-$7 & 1.441$-$7 & 1.424$-$7 & 1.424$-$7 & 1.439$-$7 & 1.567$-$7 & 1.717$-$7 & 1.861$-$7 & 1.997$-$7  \\
    2 &  75 & 2.337$-$7 & 1.884$-$7 & 1.578$-$7 & 1.373$-$7 & 1.219$-$7 & 1.114$-$7 & 8.580$-$8 & 7.859$-$8 & 7.689$-$8 & 7.717$-$8  \\
    2 &  76 & 2.474$-$6 & 2.688$-$6 & 2.894$-$6 & 3.090$-$6 & 3.266$-$6 & 3.435$-$6 & 4.119$-$6 & 4.662$-$6 & 5.128$-$6 & 5.539$-$6  \\
    2 &  77 & 1.140$-$6 & 1.138$-$6 & 1.154$-$6 & 1.181$-$6 & 1.212$-$6 & 1.247$-$6 & 1.417$-$6 & 1.574$-$6 & 1.717$-$6 & 1.845$-$6  \\
    2 &  78 & 6.218$-$8 & 3.773$-$8 & 2.460$-$8 & 1.678$-$8 & 1.201$-$8 & 8.901$-$9 & 2.857$-$9 & 1.313$-$9 & 0.731$-$9 & 0.448$-$9  \\
    2 &  79 & 9.433$-$8 & 6.978$-$8 & 5.481$-$8 & 4.511$-$8 & 3.794$-$8 & 3.337$-$8 & 2.132$-$8 & 1.690$-$8 & 1.484$-$8 & 1.369$-$8  \\
    2 &  80 & 2.801$-$7 & 2.707$-$7 & 2.698$-$7 & 2.742$-$7 & 2.806$-$7 & 2.886$-$7 & 3.305$-$7 & 3.697$-$7 & 4.049$-$7 & 4.359$-$7  \\
    2 &  81 & 4.229$-$7 & 3.940$-$7 & 3.753$-$7 & 3.609$-$7 & 3.483$-$7 & 3.408$-$7 & 3.090$-$7 & 2.895$-$7 & 2.774$-$7 & 2.677$-$7  \\
    2 &  82 & 3.821$-$6 & 3.942$-$6 & 4.033$-$6 & 4.156$-$6 & 4.196$-$6 & 4.292$-$6 & 4.543$-$6 & 4.742$-$6 & 4.967$-$6 & 5.151$-$6  \\
    2 &  83 & 8.013$-$7 & 6.047$-$7 & 4.719$-$7 & 3.792$-$7 & 3.109$-$7 & 2.611$-$7 & 1.318$-$7 & 8.342$-$8 & 6.032$-$8 & 4.727$-$8  \\
    2 &  84 & 2.782$-$6 & 3.008$-$6 & 3.225$-$6 & 3.436$-$6 & 3.628$-$6 & 3.812$-$6 & 4.562$-$6 & 5.162$-$6 & 5.679$-$6 & 6.136$-$6  \\
    2 &  85 & 1.469$-$6 & 1.516$-$6 & 1.570$-$6 & 1.609$-$6 & 1.646$-$6 & 1.683$-$6 & 1.786$-$6 & 1.848$-$6 & 1.900$-$6 & 1.945$-$6  \\
    2 &  86 & 1.307$-$4 & 1.244$-$4 & 1.235$-$4 & 1.250$-$4 & 1.283$-$4 & 1.338$-$4 & 1.670$-$4 & 2.039$-$4 & 2.404$-$4 & 2.747$-$4  \\
    2 &  87 & 1.941$-$3 & 1.973$-$3 & 2.004$-$3 & 2.043$-$3 & 2.057$-$3 & 2.097$-$3 & 2.217$-$3 & 2.328$-$3 & 2.460$-$3 & 2.558$-$3  \\
    2 &  88 & 3.779$-$4 & 3.787$-$4 & 3.810$-$4 & 3.857$-$4 & 3.877$-$4 & 3.937$-$4 & 4.152$-$4 & 4.366$-$4 & 4.610$-$4 & 4.803$-$4  \\
    2 &  89 & 2.015$-$4 & 1.587$-$4 & 1.289$-$4 & 1.073$-$4 & 9.076$-$5 & 7.875$-$5 & 4.512$-$5 & 3.163$-$5 & 2.500$-$5 & 2.082$-$5  \\
    2 &  90 & 3.390$-$3 & 3.572$-$3 & 3.718$-$3 & 3.855$-$3 & 3.944$-$3 & 4.060$-$3 & 4.435$-$3 & 4.729$-$3 & 5.027$-$3 & 5.262$-$3  \\
    2 &  91 & 1.792$-$4 & 2.024$-$4 & 2.277$-$4 & 2.532$-$4 & 2.783$-$4 & 3.054$-$4 & 4.283$-$4 & 5.394$-$4 & 6.429$-$4 & 7.394$-$4  \\
    2 &  92 & 4.043$-$4 & 3.521$-$4 & 3.173$-$4 & 2.944$-$4 & 2.768$-$4 & 2.679$-$4 & 2.513$-$4 & 2.595$-$4 & 2.751$-$4 & 2.923$-$4  \\
    2 &  93 & 8.618$-$4 & 1.042$-$3 & 1.219$-$3 & 1.384$-$3 & 1.548$-$3 & 1.716$-$3 & 2.449$-$3 & 3.090$-$3 & 3.681$-$3 & 4.228$-$3  \\
    2 &  94 & 1.811$-$4 & 2.077$-$4 & 2.365$-$4 & 2.652$-$4 & 2.935$-$4 & 3.237$-$4 & 4.603$-$4 & 5.829$-$4 & 6.968$-$4 & 8.030$-$4  \\
    2 &  95 & 4.309$-$3 & 4.870$-$3 & 5.426$-$3 & 5.926$-$3 & 6.420$-$3 & 6.914$-$3 & 9.059$-$3 & 1.092$-$2 & 1.265$-$2 & 1.423$-$2  \\
    2 &  96 & 7.202$-$4 & 5.871$-$4 & 5.035$-$4 & 4.449$-$4 & 4.056$-$4 & 3.792$-$4 & 3.253$-$4 & 3.216$-$4 & 3.319$-$4 & 3.463$-$4  \\
    2 &  97 & 4.919$-$5 & 6.610$-$5 & 8.345$-$5 & 1.005$-$4 & 1.171$-$4 & 1.344$-$4 & 2.121$-$4 & 2.809$-$4 & 3.442$-$4 & 4.032$-$4  \\
    2 &  98 & 1.895$-$4 & 2.128$-$4 & 2.412$-$4 & 2.708$-$4 & 3.025$-$4 & 3.364$-$4 & 4.986$-$4 & 6.479$-$4 & 7.878$-$4 & 9.193$-$4  \\
    2 &  99 & 1.541$-$4 & 1.233$-$4 & 1.010$-$4 & 8.368$-$5 & 7.092$-$5 & 6.014$-$5 & 3.169$-$5 & 1.938$-$5 & 1.319$-$5 & 9.693$-$6  \\
    2 & 100 & 2.797$-$4 & 3.622$-$4 & 4.487$-$4 & 5.330$-$4 & 6.183$-$4 & 7.067$-$4 & 1.109$-$3 & 1.466$-$3 & 1.796$-$3 & 2.103$-$3  \\
    2 & 101 & 1.451$-$4 & 1.956$-$4 & 2.469$-$4 & 2.973$-$4 & 3.459$-$4 & 3.966$-$4 & 6.226$-$4 & 8.221$-$4 & 1.006$-$3 & 1.177$-$3  \\
    2 & 102 & 8.443$-$6 & 6.736$-$6 & 5.556$-$6 & 4.705$-$6 & 4.073$-$6 & 3.603$-$6 & 2.404$-$6 & 1.968$-$6 & 1.793$-$6 & 1.708$-$6  \\
    2 & 103 & 3.389$-$4 & 3.252$-$4 & 3.200$-$4 & 3.201$-$4 & 3.233$-$4 & 3.286$-$4 & 3.647$-$4 & 4.024$-$4 & 4.376$-$4 & 4.704$-$4  \\
    2 & 104 & 7.945$-$4 & 8.495$-$4 & 9.051$-$4 & 9.592$-$4 & 1.013$-$3 & 1.064$-$3 & 1.282$-$3 & 1.457$-$3 & 1.606$-$3 & 1.739$-$3  \\
    2 & 105 & 4.277$-$4 & 4.385$-$4 & 4.540$-$4 & 4.716$-$4 & 4.907$-$4 & 5.097$-$4 & 5.988$-$4 & 6.745$-$4 & 7.403$-$4 & 7.999$-$4  \\
    2 & 106 & 9.861$-$5 & 7.657$-$5 & 6.100$-$5 & 4.977$-$5 & 4.116$-$5 & 3.480$-$5 & 1.754$-$5 & 1.058$-$5 & 7.066$-$6 & 5.100$-$6  \\
    2 & 107 & 6.487$-$4 & 6.689$-$4 & 6.949$-$4 & 7.247$-$4 & 7.559$-$4 & 7.872$-$4 & 9.307$-$4 & 1.052$-$3 & 1.156$-$3 & 1.251$-$3  \\
    2 & 108 & 4.038$-$4 & 3.104$-$4 & 2.447$-$4 & 1.976$-$4 & 1.625$-$4 & 1.362$-$4 & 6.662$-$5 & 3.927$-$5 & 2.562$-$5 & 1.795$-$5  \\
    2 & 109 & 1.216$-$3 & 1.387$-$3 & 1.540$-$3 & 1.679$-$3 & 1.806$-$3 & 1.923$-$3 & 2.394$-$3 & 2.751$-$3 & 3.047$-$3 & 3.309$-$3  \\
    2 & 110 & 1.003$-$3 & 1.169$-$3 & 1.314$-$3 & 1.444$-$3 & 1.561$-$3 & 1.668$-$3 & 2.094$-$3 & 2.412$-$3 & 2.675$-$3 & 2.906$-$3  \\
    2 & 111 & 3.398$-$6 & 3.260$-$6 & 3.186$-$6 & 3.180$-$6 & 3.200$-$6 & 3.252$-$6 & 3.614$-$6 & 4.006$-$6 & 4.371$-$6 & 4.716$-$6  \\
    2 & 112 & 1.288$-$5 & 1.426$-$5 & 1.570$-$5 & 1.702$-$5 & 1.836$-$5 & 1.966$-$5 & 2.552$-$5 & 3.066$-$5 & 3.540$-$5 & 3.978$-$5  \\
    2 & 113 & 3.345$-$6 & 2.710$-$6 & 2.305$-$6 & 2.021$-$6 & 1.847$-$6 & 1.716$-$6 & 1.471$-$6 & 1.456$-$6 & 1.505$-$6 & 1.571$-$6  \\
\hline	
\end{tabular}
\newpage
\begin{tabular}{rrllllllllll} \hline	  
    I &   J &      1000 &    1200   &    1400   &    1600   &     1800  &    2000   &	3000    &    4000   &    5000   &      6000  \\
\hline
    3 &   4 & 2.434$-$7 & 2.524$-$7 & 2.612$-$7 & 2.702$-$7 & 2.793$-$7 & 2.867$-$7 & 3.219$-$7 & 3.514$-$7 & 3.774$-$7 & 4.051$-$7  \\
    3 &   5 & 2.748$-$7 & 2.877$-$7 & 2.988$-$7 & 3.079$-$7 & 3.175$-$7 & 3.250$-$7 & 3.555$-$7 & 3.787$-$7 & 3.991$-$7 & 4.208$-$7  \\
    3 &   6 & 5.979$-$7 & 6.029$-$7 & 6.111$-$7 & 6.162$-$7 & 6.256$-$7 & 6.307$-$7 & 6.683$-$7 & 7.022$-$7 & 7.379$-$7 & 7.716$-$7  \\
    3 &   7 & 3.712$-$7 & 4.173$-$7 & 4.636$-$7 & 5.128$-$7 & 5.622$-$7 & 6.039$-$7 & 8.027$-$7 & 9.874$-$7 & 1.156$-$6 & 1.320$-$6  \\
    3 &   8 & 1.561$-$7 & 1.329$-$7 & 1.172$-$7 & 1.065$-$7 & 9.907$-$8 & 9.407$-$8 & 8.289$-$8 & 8.127$-$8 & 8.264$-$8 & 8.504$-$8  \\
    3 &   9 & 5.682$-$7 & 6.166$-$7 & 6.667$-$7 & 7.190$-$7 & 7.724$-$7 & 8.179$-$7 & 1.037$-$6 & 1.242$-$6 & 1.430$-$6 & 1.611$-$6  \\
    3 &  10 & 1.404$-$7 & 1.128$-$7 & 9.272$-$8 & 7.855$-$8 & 6.741$-$8 & 5.966$-$8 & 3.821$-$8 & 3.045$-$8 & 2.722$-$8 & 2.582$-$8  \\
    3 &  11 & 4.125$-$7 & 4.922$-$7 & 5.692$-$7 & 6.469$-$7 & 7.223$-$7 & 7.902$-$7 & 1.094$-$6 & 1.369$-$6 & 1.618$-$6 & 1.854$-$6  \\
    3 &  12 & 1.720$-$7 & 1.422$-$7 & 1.230$-$7 & 1.115$-$7 & 1.038$-$7 & 9.876$-$8 & 9.111$-$8 & 9.309$-$8 & 9.739$-$8 & 1.024$-$7  \\
    3 &  13 & 2.111$-$6 & 2.335$-$6 & 2.551$-$6 & 2.771$-$6 & 2.992$-$6 & 3.182$-$6 & 4.078$-$6 & 4.907$-$6 & 5.678$-$6 & 6.407$-$6  \\
    3 &  14 & 9.870$-$7 & 9.036$-$7 & 8.339$-$7 & 7.916$-$7 & 7.484$-$7 & 7.196$-$7 & 6.251$-$7 & 5.842$-$7 & 5.771$-$7 & 5.669$-$7  \\	
    3 &  15 & 1.741$-$7 & 1.527$-$7 & 1.401$-$7 & 1.325$-$7 & 1.283$-$7 & 1.250$-$7 & 1.236$-$7 & 1.290$-$7 & 1.362$-$7 & 1.443$-$7  \\
    3 &  16 & 2.260$-$7 & 2.037$-$7 & 1.920$-$7 & 1.860$-$7 & 1.833$-$7 & 1.820$-$7 & 1.887$-$7 & 2.007$-$7 & 2.132$-$7 & 2.267$-$7  \\
    3 &  17 & 8.645$-$7 & 8.777$-$7 & 8.892$-$7 & 8.962$-$7 & 9.083$-$7 & 9.120$-$7 & 9.469$-$7 & 9.793$-$7 & 1.020$-$6 & 1.053$-$6  \\
    3 &  18 & 9.389$-$8 & 7.252$-$8 & 6.039$-$8 & 5.312$-$8 & 4.839$-$8 & 4.529$-$8 & 3.777$-$8 & 3.505$-$8 & 3.385$-$8 & 3.322$-$8  \\
    3 &  19 & 9.502$-$7 & 9.789$-$7 & 1.011$-$6 & 1.049$-$6 & 1.086$-$6 & 1.121$-$6 & 1.282$-$6 & 1.420$-$6 & 1.543$-$6 & 1.659$-$6  \\
    3 &  20 & 1.223$-$7 & 8.849$-$8 & 6.767$-$8 & 5.399$-$8 & 4.458$-$8 & 3.761$-$8 & 2.072$-$8 & 1.423$-$8 & 1.112$-$8 & 9.255$-$9  \\
    3 &  21 & 8.107$-$8 & 5.416$-$8 & 3.995$-$8 & 3.193$-$8 & 2.714$-$8 & 2.418$-$8 & 1.916$-$8 & 1.879$-$8 & 1.935$-$8 & 2.016$-$8  \\
    3 &  22 & 2.049$-$6 & 2.085$-$6 & 2.114$-$6 & 2.134$-$6 & 2.163$-$6 & 2.175$-$6 & 2.263$-$6 & 2.344$-$6 & 2.444$-$6 & 2.524$-$6  \\
    3 &  23 & 6.924$-$7 & 6.915$-$7 & 6.938$-$7 & 7.041$-$7 & 7.098$-$7 & 7.178$-$7 & 7.482$-$7 & 7.743$-$7 & 8.005$-$7 & 8.249$-$7  \\
    3 &  24 & 1.476$-$6 & 1.460$-$6 & 1.442$-$6 & 1.420$-$6 & 1.403$-$6 & 1.385$-$6 & 1.294$-$6 & 1.225$-$6 & 1.181$-$6 & 1.150$-$6  \\
    3 &  25 & 2.284$-$6 & 2.384$-$6 & 2.481$-$6 & 2.589$-$6 & 2.681$-$6 & 2.769$-$6 & 3.122$-$6 & 3.395$-$6 & 3.629$-$6 & 3.841$-$6  \\
    3 &  26 & 2.473$-$7 & 2.038$-$7 & 1.713$-$7 & 1.492$-$7 & 1.313$-$7 & 1.178$-$7 & 8.359$-$8 & 7.127$-$8 & 6.696$-$8 & 6.611$-$8  \\
    3 &  27 & 3.614$-$7 & 3.642$-$7 & 3.662$-$7 & 3.699$-$7 & 3.733$-$7 & 3.763$-$7 & 3.937$-$7 & 4.110$-$7 & 4.309$-$7 & 4.484$-$7  \\
    3 &  28 & 7.277$-$7 & 7.554$-$7 & 7.844$-$7 & 8.177$-$7 & 8.481$-$7 & 8.750$-$7 & 9.998$-$7 & 1.106$-$6 & 1.200$-$6 & 1.290$-$6  \\
    3 &  29 & 9.126$-$7 & 9.675$-$7 & 1.019$-$6 & 1.073$-$6 & 1.119$-$6 & 1.160$-$6 & 1.339$-$6 & 1.486$-$6 & 1.614$-$6 & 1.736$-$6  \\
    3 &  30 & 1.284$-$7 & 1.111$-$7 & 9.847$-$8 & 9.081$-$8 & 8.573$-$8 & 8.250$-$8 & 7.992$-$8 & 8.656$-$8 & 9.566$-$8 & 1.053$-$7  \\
    3 &  31 & 4.661$-$7 & 5.044$-$7 & 5.510$-$7 & 6.026$-$7 & 6.569$-$7 & 7.135$-$7 & 9.734$-$7 & 1.217$-$6 & 1.443$-$6 & 1.654$-$6  \\
    3 &  32 & 4.208$-$7 & 3.640$-$7 & 3.253$-$7 & 3.007$-$7 & 2.820$-$7 & 2.691$-$7 & 2.434$-$7 & 2.434$-$7 & 2.512$-$7 & 2.617$-$7  \\
    3 &  33 & 3.590$-$6 & 4.181$-$6 & 4.759$-$6 & 5.320$-$6 & 5.866$-$6 & 6.419$-$6 & 8.815$-$6 & 1.098$-$5 & 1.297$-$5 & 1.481$-$5  \\
    3 &  34 & 1.645$-$4 & 1.312$-$4 & 1.076$-$4 & 8.921$-$5 & 7.512$-$5 & 6.413$-$5 & 3.307$-$5 & 2.006$-$5 & 1.354$-$5 & 9.459$-$6  \\
    3 &  35 & 4.200$-$4 & 5.318$-$4 & 6.413$-$4 & 7.449$-$4 & 8.467$-$4 & 9.490$-$4 & 1.389$-$3 & 1.777$-$3 & 2.129$-$3 & 2.453$-$3  \\
    3 &  36 & 1.813$-$4 & 2.313$-$4 & 2.803$-$4 & 3.264$-$4 & 3.719$-$4 & 4.176$-$4 & 6.138$-$4 & 7.866$-$4 & 9.431$-$4 & 1.088$-$3  \\
    3 &  37 & 1.038$-$4 & 8.291$-$5 & 6.833$-$5 & 5.688$-$5 & 4.819$-$5 & 4.141$-$5 & 2.245$-$5 & 1.470$-$5 & 1.098$-$5 & 8.749$-$6  \\
    3 &  38 & 4.404$-$6 & 3.475$-$6 & 2.837$-$6 & 2.338$-$6 & 1.964$-$6 & 1.671$-$6 & 8.568$-$7 & 5.200$-$7 & 3.523$-$7 & 2.469$-$7  \\
    3 &  39 & 4.390$-$6 & 3.764$-$6 & 3.386$-$6 & 3.129$-$6 & 2.975$-$6 & 2.897$-$6 & 2.947$-$6 & 3.318$-$6 & 3.771$-$6 & 4.228$-$6  \\
    3 &  40 & 4.290$-$5 & 5.437$-$5 & 6.553$-$5 & 7.595$-$5 & 8.623$-$5 & 9.653$-$5 & 1.407$-$4 & 1.795$-$4 & 2.146$-$4 & 2.471$-$4  \\
    3 &  41 & 3.818$-$7 & 2.987$-$7 & 2.446$-$7 & 2.110$-$7 & 1.866$-$7 & 1.699$-$7 & 1.339$-$7 & 1.264$-$7 & 1.270$-$7 & 1.304$-$7  \\
    3 &  42 & 2.367$-$4 & 1.909$-$4 & 1.586$-$4 & 1.362$-$4 & 1.175$-$4 & 1.041$-$4 & 6.736$-$5 & 5.312$-$5 & 4.705$-$5 & 4.456$-$5  \\
    3 &  43 & 1.769$-$5 & 1.783$-$5 & 1.850$-$5 & 1.935$-$5 & 2.042$-$5 & 2.167$-$5 & 2.802$-$5 & 3.441$-$5 & 4.052$-$5 & 4.627$-$5  \\
    3 &  44 & 5.206$-$4 & 5.136$-$4 & 5.173$-$4 & 5.271$-$4 & 5.379$-$4 & 5.493$-$4 & 6.143$-$4 & 6.778$-$4 & 7.364$-$4 & 7.926$-$4  \\
    3 &  45 & 1.524$-$6 & 1.206$-$6 & 9.942$-$7 & 8.341$-$7 & 7.155$-$7 & 6.254$-$7 & 3.865$-$7 & 2.974$-$7 & 2.591$-$7 & 2.390$-$7  \\
    3 &  46 & 3.497$-$6 & 3.663$-$6 & 3.913$-$6 & 4.194$-$6 & 4.494$-$6 & 4.819$-$6 & 6.315$-$6 & 7.711$-$6 & 9.000$-$6 & 1.018$-$5  \\
    3 &  47 & 1.152$-$4 & 9.093$-$5 & 7.372$-$5 & 6.164$-$5 & 5.139$-$5 & 4.407$-$5 & 2.319$-$5 & 1.431$-$5 & 9.882$-$6 & 7.386$-$6  \\
    3 &  48 & 3.454$-$4 & 3.391$-$4 & 3.402$-$4 & 3.457$-$4 & 3.516$-$4 & 3.587$-$4 & 3.991$-$4 & 4.396$-$4 & 4.773$-$4 & 5.131$-$4  \\
    3 &  49 & 2.402$-$7 & 2.294$-$7 & 2.228$-$7 & 2.189$-$7 & 2.150$-$7 & 2.130$-$7 & 2.095$-$7 & 2.130$-$7 & 2.208$-$7 & 2.271$-$7  \\
    3 &  50 & 1.465$-$6 & 1.415$-$6 & 1.400$-$6 & 1.409$-$6 & 1.420$-$6 & 1.439$-$6 & 1.571$-$6 & 1.715$-$6 & 1.854$-$6 & 1.989$-$6  \\
\hline	
\end{tabular}
\newpage
\begin{tabular}{rrllllllllll} \hline	  
    I &   J &      1000 &    1200   &    1400   &    1600   &     1800  &    2000   &	3000    &    4000   &    5000   &      6000  \\
\hline
    3 &  51 & 1.246$-$6 & 1.124$-$6 & 1.055$-$6 & 1.017$-$6 & 9.906$-$7 & 9.772$-$7 & 9.866$-$7 & 1.042$-$6 & 1.109$-$6 & 1.178$-$6  \\
    3 &  52 & 1.715$-$7 & 1.212$-$7 & 9.163$-$8 & 7.366$-$8 & 6.123$-$8 & 5.302$-$8 & 3.477$-$8 & 2.958$-$8 & 2.786$-$8 & 2.722$-$8  \\
    3 &  53 & 4.042$-$6 & 3.500$-$6 & 3.141$-$6 & 2.890$-$6 & 2.685$-$6 & 2.539$-$6 & 2.150$-$6 & 2.016$-$6 & 1.992$-$6 & 1.980$-$6  \\
    3 &  54 & 3.064$-$6 & 2.996$-$6 & 3.002$-$6 & 3.046$-$6 & 3.099$-$6 & 3.159$-$6 & 3.503$-$6 & 3.845$-$6 & 4.163$-$6 & 4.466$-$6  \\	
    3 &  55 & 6.397$-$6 & 6.062$-$6 & 5.904$-$6 & 5.850$-$6 & 5.818$-$6 & 5.836$-$6 & 6.118$-$6 & 6.522$-$6 & 6.943$-$6 & 7.362$-$6  \\
    3 &  56 & 1.677$-$6 & 1.309$-$6 & 1.055$-$6 & 8.721$-$7 & 7.242$-$7 & 6.163$-$7 & 3.134$-$7 & 1.888$-$7 & 1.277$-$7 & 9.059$-$8  \\
    3 &  57 & 3.782$-$7 & 3.484$-$7 & 3.322$-$7 & 3.252$-$7 & 3.207$-$7 & 3.198$-$7 & 3.342$-$7 & 3.596$-$7 & 3.864$-$7 & 4.133$-$7  \\
    3 &  58 & 7.035$-$7 & 5.916$-$7 & 5.128$-$7 & 4.620$-$7 & 4.186$-$7 & 3.904$-$7 & 3.159$-$7 & 2.947$-$7 & 2.914$-$7 & 2.955$-$7  \\
    3 &  59 & 1.189$-$7 & 7.837$-$8 & 5.710$-$8 & 4.503$-$8 & 3.787$-$8 & 3.335$-$8 & 2.614$-$8 & 2.573$-$8 & 2.660$-$8 & 2.772$-$8  \\
    3 &  60 & 1.692$-$6 & 1.702$-$6 & 1.744$-$6 & 1.809$-$6 & 1.876$-$6 & 1.941$-$6 & 2.270$-$6 & 2.567$-$6 & 2.829$-$6 & 3.074$-$6  \\
    3 &  61 & 7.084$-$7 & 5.469$-$7 & 4.390$-$7 & 3.630$-$7 & 3.029$-$7 & 2.594$-$7 & 1.394$-$7 & 9.051$-$8 & 6.646$-$8 & 5.179$-$8  \\
    3 &  62 & 1.288$-$6 & 1.177$-$6 & 1.114$-$6 & 1.076$-$6 & 1.039$-$6 & 1.016$-$6 & 9.371$-$7 & 8.921$-$7 & 8.645$-$7 & 8.390$-$7  \\
    3 &  63 & 1.975$-$5 & 2.037$-$5 & 2.096$-$5 & 2.152$-$5 & 2.190$-$5 & 2.234$-$5 & 2.392$-$5 & 2.530$-$5 & 2.678$-$5 & 2.789$-$5  \\
    3 &  64 & 3.851$-$4 & 2.966$-$4 & 2.373$-$4 & 1.944$-$4 & 1.604$-$4 & 1.356$-$4 & 6.734$-$5 & 3.984$-$5 & 2.651$-$5 & 1.853$-$5  \\
    3 &  65 & 2.571$-$4 & 2.092$-$4 & 1.761$-$4 & 1.540$-$4 & 1.352$-$4 & 1.228$-$4 & 8.923$-$5 & 7.825$-$5 & 7.493$-$5 & 7.484$-$5  \\
    3 &  66 & 3.377$-$4 & 3.246$-$4 & 3.217$-$4 & 3.248$-$4 & 3.294$-$4 & 3.359$-$4 & 3.768$-$4 & 4.193$-$4 & 4.589$-$4 & 4.963$-$4  \\
    3 &  67 & 1.839$-$3 & 1.877$-$3 & 1.917$-$3 & 1.960$-$3 & 1.985$-$3 & 2.021$-$3 & 2.149$-$3 & 2.267$-$3 & 2.397$-$3 & 2.497$-$3  \\
    3 &  68 & 2.985$-$4 & 2.380$-$4 & 1.984$-$4 & 1.703$-$4 & 1.484$-$4 & 1.330$-$4 & 9.334$-$5 & 8.084$-$5 & 7.726$-$5 & 7.691$-$5  \\
    3 &  69 & 2.428$-$4 & 2.296$-$4 & 2.243$-$4 & 2.240$-$4 & 2.247$-$4 & 2.279$-$4 & 2.502$-$4 & 2.760$-$4 & 3.007$-$4 & 3.244$-$4  \\
    3 &  70 & 5.823$-$3 & 6.023$-$3 & 6.210$-$3 & 6.390$-$3 & 6.503$-$3 & 6.647$-$3 & 7.138$-$3 & 7.558$-$3 & 8.010$-$3 & 8.349$-$3  \\
    3 &  71 & 3.735$-$4 & 3.083$-$4 & 2.642$-$4 & 2.352$-$4 & 2.147$-$4 & 2.009$-$4 & 1.755$-$4 & 1.792$-$4 & 1.916$-$4 & 2.067$-$4  \\
    3 &  72 & 7.785$-$4 & 7.271$-$4 & 7.038$-$4 & 6.977$-$4 & 7.020$-$4 & 7.151$-$4 & 8.167$-$4 & 9.392$-$4 & 1.063$-$3 & 1.182$-$3  \\
    3 &  73 & 5.310$-$4 & 4.817$-$4 & 4.558$-$4 & 4.442$-$4 & 4.381$-$4 & 4.383$-$4 & 4.619$-$4 & 4.979$-$4 & 5.346$-$4 & 5.701$-$4  \\
    3 &  74 & 3.344$-$4 & 2.554$-$4 & 2.030$-$4 & 1.687$-$4 & 1.435$-$4 & 1.260$-$4 & 8.530$-$5 & 7.419$-$5 & 7.149$-$5 & 7.184$-$5  \\
    3 &  75 & 4.088$-$4 & 3.649$-$4 & 3.391$-$4 & 3.252$-$4 & 3.183$-$4 & 3.171$-$4 & 3.411$-$4 & 3.840$-$4 & 4.309$-$4 & 4.771$-$4  \\
    3 &  76 & 4.365$-$4 & 3.982$-$4 & 3.783$-$4 & 3.699$-$4 & 3.655$-$4 & 3.663$-$4 & 3.873$-$4 & 4.178$-$4 & 4.488$-$4 & 4.786$-$4  \\
    3 &  77 & 3.573$-$4 & 3.024$-$4 & 2.684$-$4 & 2.484$-$4 & 2.359$-$4 & 2.298$-$4 & 2.336$-$4 & 2.586$-$4 & 2.887$-$4 & 3.190$-$4  \\
    3 &  78 & 5.931$-$4 & 4.371$-$4 & 3.343$-$4 & 2.626$-$4 & 2.105$-$4 & 1.723$-$4 & 7.719$-$5 & 4.305$-$5 & 2.744$-$5 & 1.888$-$5  \\
    3 &  79 & 3.775$-$4 & 3.414$-$4 & 3.239$-$4 & 3.178$-$4 & 3.157$-$4 & 3.181$-$4 & 3.438$-$4 & 3.749$-$4 & 4.049$-$4 & 4.332$-$4  \\
    3 &  80 & 3.987$-$4 & 2.955$-$4 & 2.263$-$4 & 1.789$-$4 & 1.443$-$4 & 1.188$-$4 & 5.538$-$5 & 3.265$-$5 & 2.226$-$5 & 1.674$-$5  \\
    3 &  81 & 4.215$-$3 & 4.751$-$3 & 5.267$-$3 & 5.740$-$3 & 6.196$-$3 & 6.654$-$3 & 8.619$-$3 & 1.034$-$2 & 1.192$-$2 & 1.337$-$2  \\
    3 &  82 & 7.658$-$3 & 8.824$-$3 & 9.909$-$3 & 1.089$-$2 & 1.182$-$2 & 1.274$-$2 & 1.664$-$2 & 2.001$-$2 & 2.310$-$2 & 2.593$-$2  \\
    3 &  83 & 4.476$-$4 & 3.335$-$4 & 2.591$-$4 & 2.080$-$4 & 1.710$-$4 & 1.443$-$4 & 7.971$-$5 & 5.842$-$5 & 4.997$-$5 & 4.629$-$5  \\
    3 &  84 & 3.289$-$4 & 2.794$-$4 & 2.500$-$4 & 2.333$-$4 & 2.223$-$4 & 2.166$-$4 & 2.133$-$4 & 2.246$-$4 & 2.388$-$4 & 2.534$-$4  \\
    3 &  85 & 1.161$-$2 & 1.337$-$2 & 1.501$-$2 & 1.649$-$2 & 1.789$-$2 & 1.929$-$2 & 2.522$-$2 & 3.034$-$2 & 3.504$-$2 & 3.934$-$2  \\
    3 &  86 & 1.179$-$6 & 1.153$-$6 & 1.148$-$6 & 1.153$-$6 & 1.161$-$6 & 1.172$-$6 & 1.248$-$6 & 1.330$-$6 & 1.414$-$6 & 1.486$-$6  \\
    3 &  87 & 1.422$-$6 & 1.335$-$6 & 1.306$-$6 & 1.295$-$6 & 1.308$-$6 & 1.338$-$6 & 1.564$-$6 & 1.837$-$6 & 2.117$-$6 & 2.385$-$6  \\
    3 &  88 & 6.263$-$5 & 5.783$-$5 & 5.616$-$5 & 5.564$-$5 & 5.632$-$5 & 5.798$-$5 & 7.052$-$5 & 8.560$-$5 & 1.008$-$4 & 1.153$-$4  \\
    3 &  89 & 1.334$-$4 & 1.344$-$4 & 1.411$-$4 & 1.494$-$4 & 1.595$-$4 & 1.721$-$4 & 2.357$-$4 & 2.984$-$4 & 3.579$-$4 & 4.134$-$4  \\
    3 &  90 & 6.425$-$5 & 5.598$-$5 & 5.117$-$5 & 4.771$-$5 & 4.565$-$5 & 4.453$-$5 & 4.541$-$5 & 5.067$-$5 & 5.711$-$5 & 6.359$-$5  \\
    3 &  91 & 2.842$-$4 & 2.672$-$4 & 2.583$-$4 & 2.503$-$4 & 2.452$-$4 & 2.422$-$4 & 2.379$-$4 & 2.422$-$4 & 2.520$-$4 & 2.593$-$4  \\
    3 &  92 & 8.111$-$6 & 9.163$-$6 & 1.051$-$5 & 1.190$-$5 & 1.336$-$5 & 1.501$-$5 & 2.268$-$5 & 2.975$-$5 & 3.633$-$5 & 4.244$-$5  \\
    3 &  93 & 2.233$-$4 & 1.753$-$4 & 1.433$-$4 & 1.177$-$4 & 9.867$-$5 & 8.422$-$5 & 4.375$-$5 & 2.689$-$5 & 1.892$-$5 & 1.406$-$5  \\
    3 &  94 & 3.735$-$3 & 3.871$-$3 & 4.013$-$3 & 4.108$-$3 & 4.199$-$3 & 4.285$-$3 & 4.618$-$3 & 4.890$-$3 & 5.183$-$3 & 5.392$-$3  \\
    3 &  95 & 8.436$-$6 & 8.490$-$6 & 8.700$-$6 & 8.953$-$6 & 9.255$-$6 & 9.560$-$6 & 1.101$-$5 & 1.227$-$5 & 1.338$-$5 & 1.438$-$5  \\	
\hline	
\end{tabular}
\newpage
\begin{tabular}{rrllllllllll} \hline	  
    I &   J &      1000 &    1200   &    1400   &    1600   &     1800  &    2000   &	3000    &    4000   &    5000   &      6000  \\
\hline
    3 &  96 & 6.534$-$6 & 6.501$-$6 & 6.583$-$6 & 6.739$-$6 & 6.930$-$6 & 7.148$-$6 & 8.245$-$6 & 9.246$-$6 & 1.014$-$5 & 1.094$-$5  \\
    3 &  97 & 1.642$-$3 & 1.676$-$3 & 1.715$-$3 & 1.744$-$3 & 1.776$-$3 & 1.805$-$3 & 1.935$-$3 & 2.048$-$3 & 2.168$-$3 & 2.261$-$3  \\
    3 &  98 & 2.388$-$4 & 2.009$-$4 & 1.763$-$4 & 1.579$-$4 & 1.452$-$4 & 1.365$-$4 & 1.171$-$4 & 1.152$-$4 & 1.191$-$4 & 1.247$-$4  \\
    3 &  99 & 3.301$-$4 & 3.011$-$4 & 2.866$-$4 & 2.773$-$4 & 2.747$-$4 & 2.743$-$4 & 2.925$-$4 & 3.205$-$4 & 3.499$-$4 & 3.777$-$4  \\
    3 & 100 & 2.342$-$4 & 1.931$-$4 & 1.657$-$4 & 1.451$-$4 & 1.306$-$4 & 1.202$-$4 & 9.493$-$5 & 8.893$-$5 & 8.952$-$5 & 9.220$-$5  \\
    3 & 101 & 4.508$-$4 & 4.593$-$4 & 4.697$-$4 & 4.772$-$4 & 4.858$-$4 & 4.936$-$4 & 5.294$-$4 & 5.604$-$4 & 5.935$-$4 & 6.194$-$4  \\
    3 & 102 & 1.081$-$3 & 1.298$-$3 & 1.514$-$3 & 1.715$-$3 & 1.910$-$3 & 2.106$-$3 & 2.975$-$3 & 3.732$-$3 & 4.425$-$3 & 5.067$-$3  \\
    3 & 103 & 9.161$-$4 & 9.849$-$4 & 1.064$-$3 & 1.141$-$3 & 1.219$-$3 & 1.302$-$3 & 1.678$-$3 & 2.017$-$3 & 2.332$-$3 & 2.623$-$3  \\
    3 & 104 & 3.303$-$4 & 2.476$-$4 & 1.914$-$4 & 1.513$-$4 & 1.230$-$4 & 1.022$-$4 & 4.995$-$5 & 3.165$-$5 & 2.367$-$5 & 1.953$-$5  \\
    3 & 105 & 4.903$-$3 & 5.699$-$3 & 6.463$-$3 & 7.148$-$3 & 7.796$-$3 & 8.445$-$3 & 1.123$-$2 & 1.363$-$2 & 1.582$-$2 & 1.783$-$2  \\
    3 & 106 & 2.294$-$3 & 2.640$-$3 & 2.972$-$3 & 3.265$-$3 & 3.547$-$3 & 3.822$-$3 & 5.010$-$3 & 6.031$-$3 & 6.964$-$3 & 7.820$-$3  \\
    3 & 107 & 4.211$-$4 & 3.323$-$4 & 2.735$-$4 & 2.320$-$4 & 2.035$-$4 & 1.832$-$4 & 1.360$-$4 & 1.242$-$4 & 1.227$-$4 & 1.249$-$4  \\
    3 & 108 & 3.867$-$4 & 3.437$-$4 & 3.207$-$4 & 3.082$-$4 & 3.034$-$4 & 3.023$-$4 & 3.193$-$4 & 3.453$-$4 & 3.718$-$4 & 3.970$-$4  \\
    3 & 109 & 3.579$-$4 & 2.890$-$4 & 2.446$-$4 & 2.142$-$4 & 1.943$-$4 & 1.805$-$4 & 1.528$-$4 & 1.504$-$4 & 1.547$-$4 & 1.612$-$4  \\
    3 & 110 & 3.647$-$4 & 2.793$-$4 & 2.218$-$4 & 1.803$-$4 & 1.511$-$4 & 1.299$-$4 & 7.707$-$5 & 6.029$-$5 & 5.444$-$5 & 5.275$-$5  \\
    3 & 111 & 3.707$-$4 & 4.305$-$4 & 5.027$-$4 & 5.781$-$4 & 6.540$-$4 & 7.359$-$4 & 1.123$-$3 & 1.477$-$3 & 1.805$-$3 & 2.114$-$3  \\
    3 & 112 & 9.909$-$4 & 1.046$-$3 & 1.105$-$3 & 1.167$-$3 & 1.227$-$3 & 1.286$-$3 & 1.544$-$3 & 1.753$-$3 & 1.930$-$3 & 2.091$-$3  \\
    3 & 113 & 1.449$-$3 & 1.532$-$3 & 1.621$-$3 & 1.715$-$3 & 1.806$-$3 & 1.896$-$3 & 2.285$-$3 & 2.600$-$3 & 2.867$-$3 & 3.109$-$3  \\
\hline				  				        				   
\end{tabular}										        				   
\begin {flushleft}									        				   
\begin{tabbing} 									      
aaaaaaaaaaaaaaaaaaaaaaaaaaaaaaaaaaaa\= \kill						      		      
		      		      
\end{tabbing}										      
\end {flushleft}									      
%\end{document}

\newpage

\setcounter{figure} {0}
\begin{figure*}
\includegraphics[angle=-90,width=0.90\textwidth]{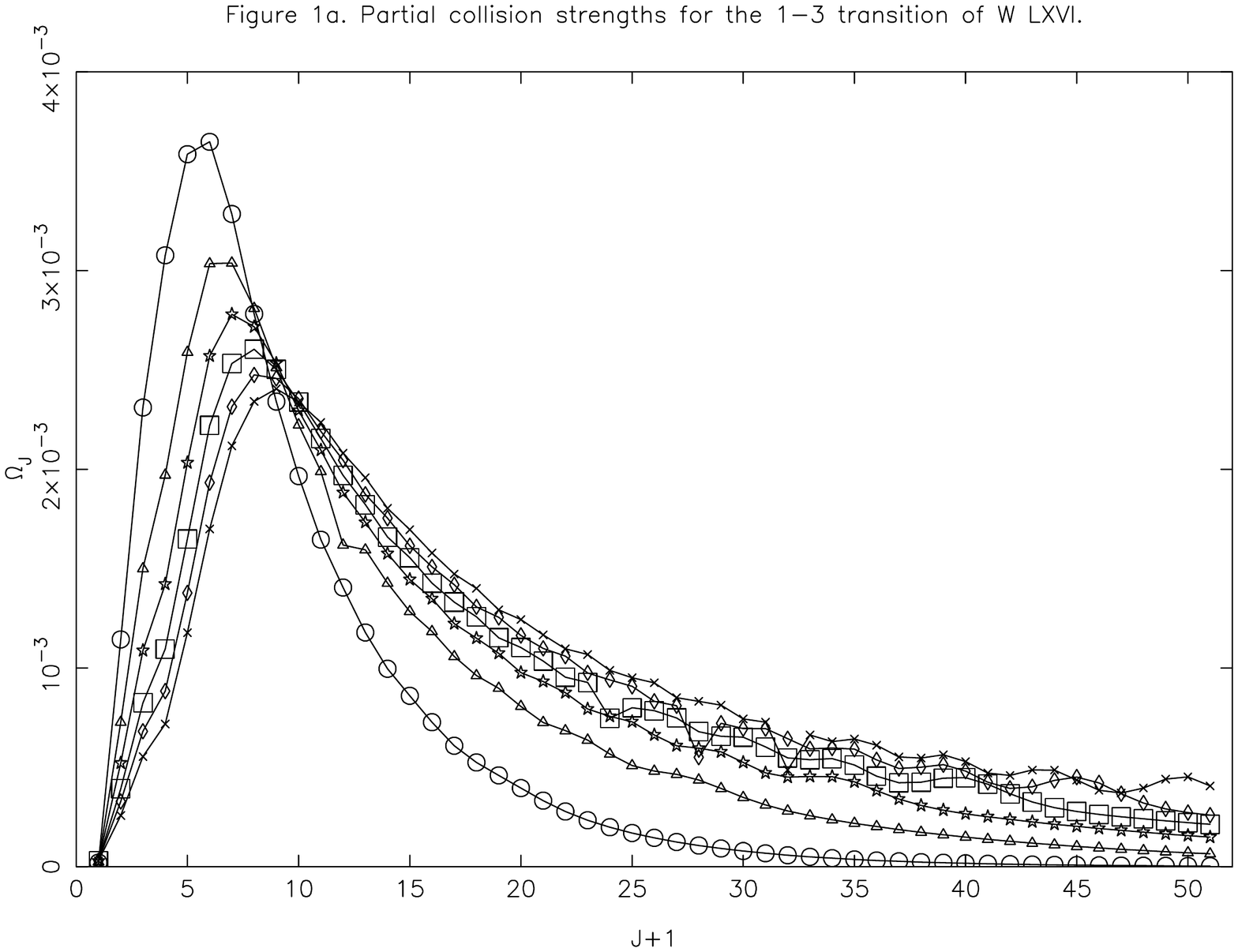}
\caption{a. Partial collision strengths for the 1-3 (2s$^2$2p$^5$ $^2$P$^o_{3/2}$ - 2s2p$^6$ $^2$S$_{1/2}$) transition of W LXVI,
at six energies of: 1000 Ryd (circles), 2000 Ryd (triangles), 3000 Ryd (stars), 4000 Ryd (squares), 5000 Ryd (diamonds), and 6000 Ryd (crosses).}
\end{figure*}

\newpage
\setcounter{figure} {0}
\begin{figure*}
\includegraphics[angle=-90,width=0.90\textwidth]{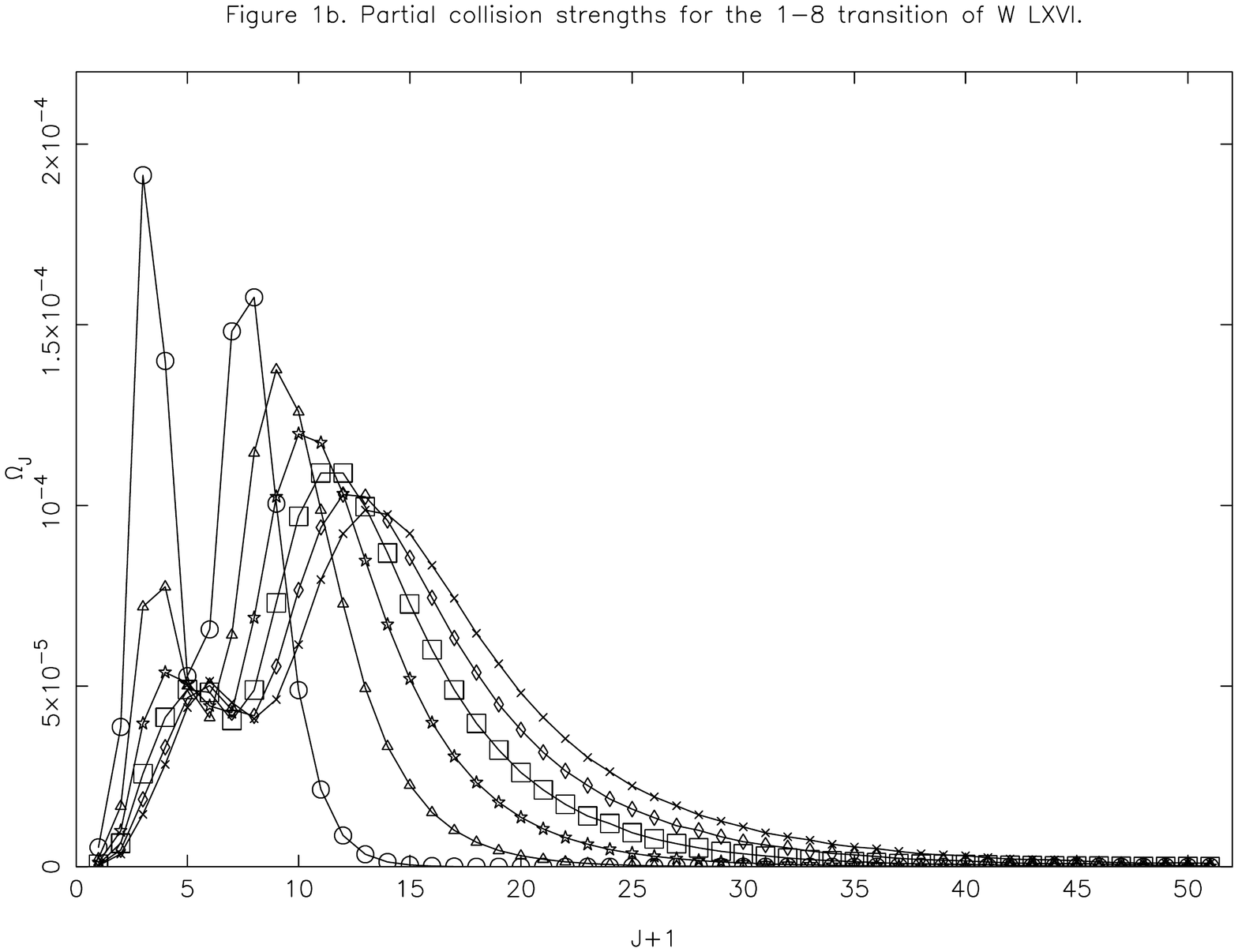}
\caption{b. Partial collision strengths for the 1-8 (2s$^2$2p$^5$ $^2$P$^o_{3/2}$ - 2s$^2$2p$^4$3p $^2$D$^o_{5/2}$) transition of W LXVI,
at six energies of: 1000 Ryd (circles), 2000 Ryd (triangles), 3000 Ryd (stars), 4000 Ryd (squares), 5000 Ryd (diamonds), and 6000 Ryd (crosses).}
\end{figure*}

\newpage
\setcounter{figure} {0}
\begin{figure*}
\includegraphics[angle=-90,width=0.90\textwidth]{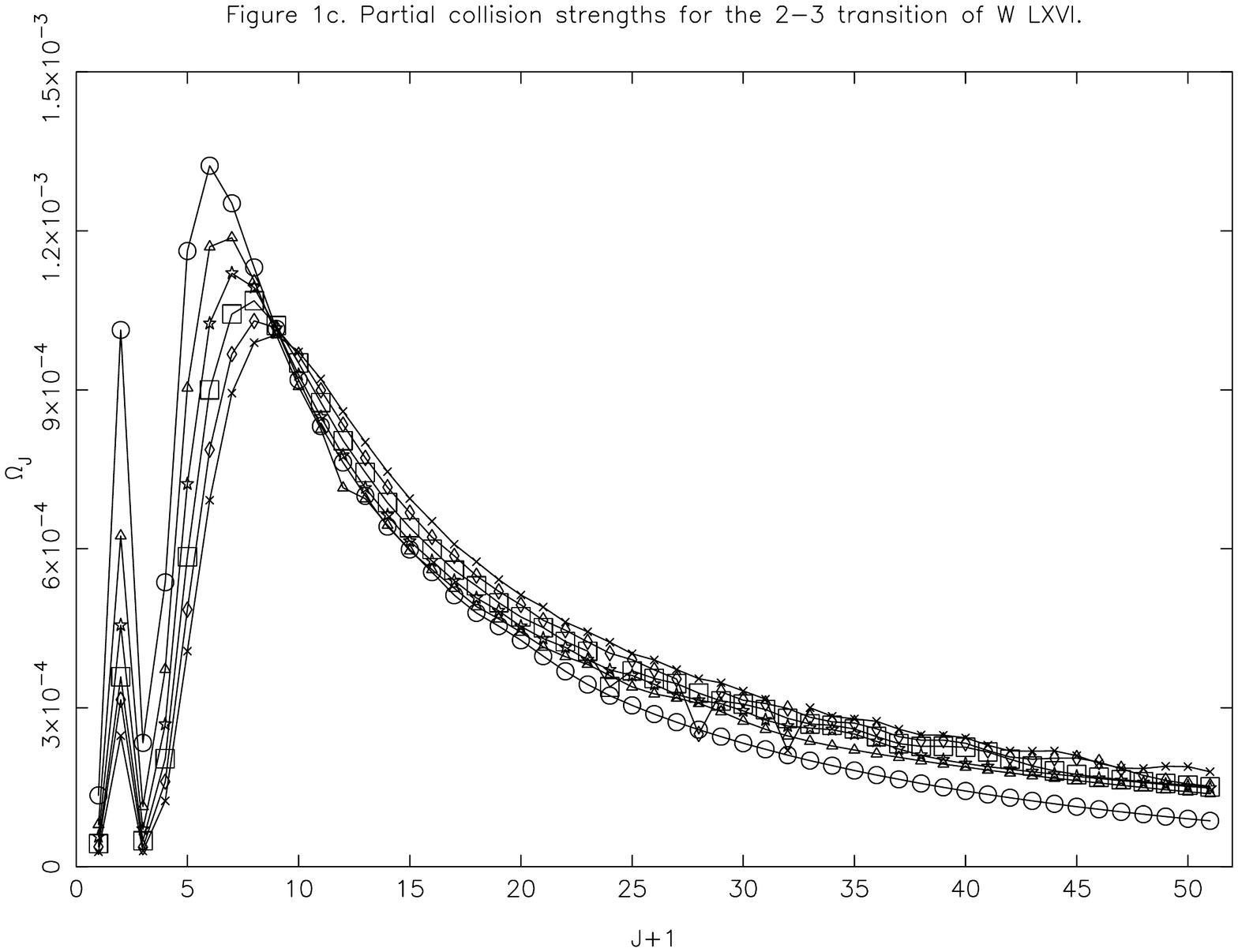}
\caption{c. Partial collision strengths for the 2-3 (2s$^2$2p$^5$ $^2$P$^o_{1/2}$ - 2s2p$^6$ $^2$S$_{1/2}$) transition of W LXVI,
at six energies of: 1000 Ryd (circles), 2000 Ryd (triangles), 3000 Ryd (stars), 4000 Ryd (squares), 5000 Ryd (diamonds), and 6000 Ryd (crosses).}
\end{figure*}

\newpage
\setcounter{figure} {1}
\begin{figure*}
\includegraphics[angle=-90,width=0.90\textwidth]{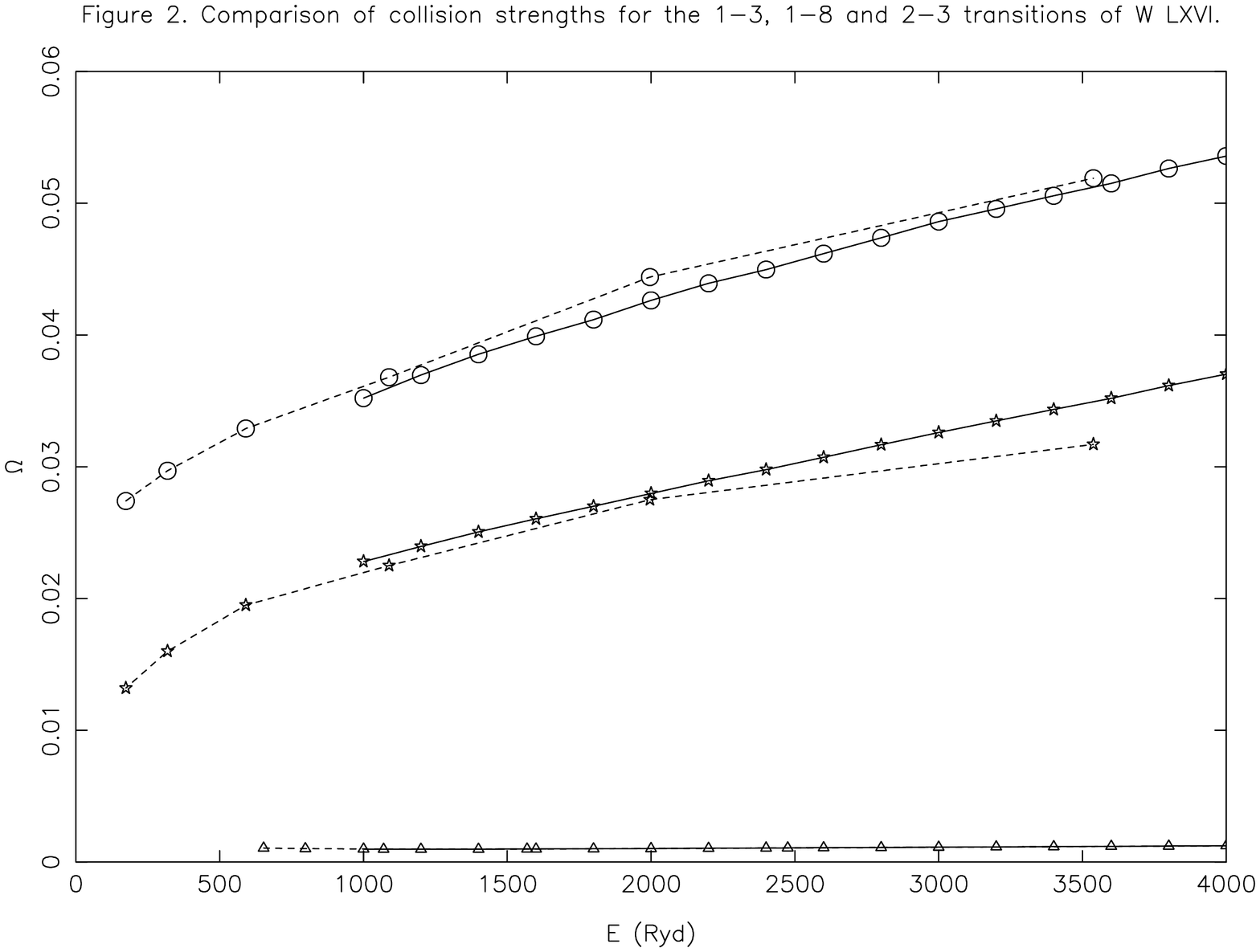}
\caption{Comparison of collision strengths for the 1-3 (2s$^2$2p$^5$ $^2$P$^o_{3/2}$ - 2s2p$^6$ $^2$S$_{1/2}$: circles), 1-8 (2s$^2$2p$^5$ $^2$P$^o_{3/2}$ - 2s$^2$2p$^4$3p $^2$D$^o_{5/2}$: triangles) and 2-3 (2s$^2$2p$^5$ $^2$P$^o_{1/2}$ - 2s2p$^6$ $^2$S$_{1/2}$: stars) transitions of W LXVI. Continuous curves: present results, broken curves: results of Sampson {\em et al.}$^{\cite{zs1}}$}
\end{figure*}

\newpage
\setcounter{figure} {2}
\begin{figure*}
\includegraphics[angle=-90,width=0.90\textwidth]{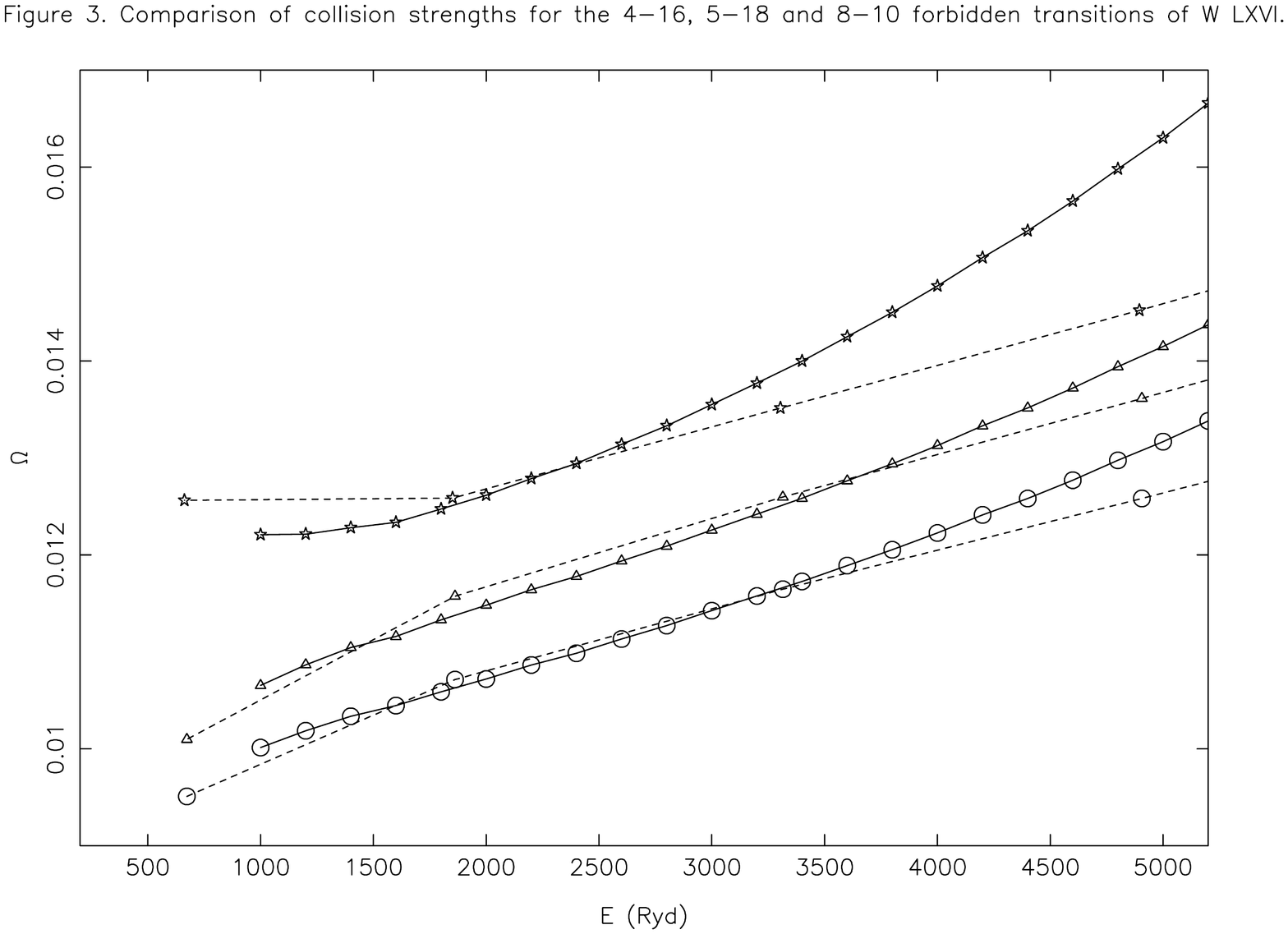}
\caption{Comparison of collision strengths for the 4--16 (2s$^2$2p$^4$3s~$^4$P$_{5/2}$ -- 2s$^2$2p$^4$3d~$^4$D$_{5/2}$: circles), 5--18 (2s$^2$2p$^4$3s~$^2$P$_{3/2}$ -- 2s$^2$2p$^4$3d~$^2$F$_{7/2}$: triangles) and 8--10 (2s$^2$2p$^4$3p~$^2$D$^o_{5/2}$ -- 2s$^2$2p$^4$3p~$^4$P$^o_{5/2}$: stars) transitions of W~LXVI. Continuous curves: present results from {\sc darc}, broken curves: present results from {\sc fac}.}
\end{figure*}

\newpage
\setcounter{figure} {3}
\begin{figure*}
\includegraphics[angle=-90,width=0.90\textwidth]{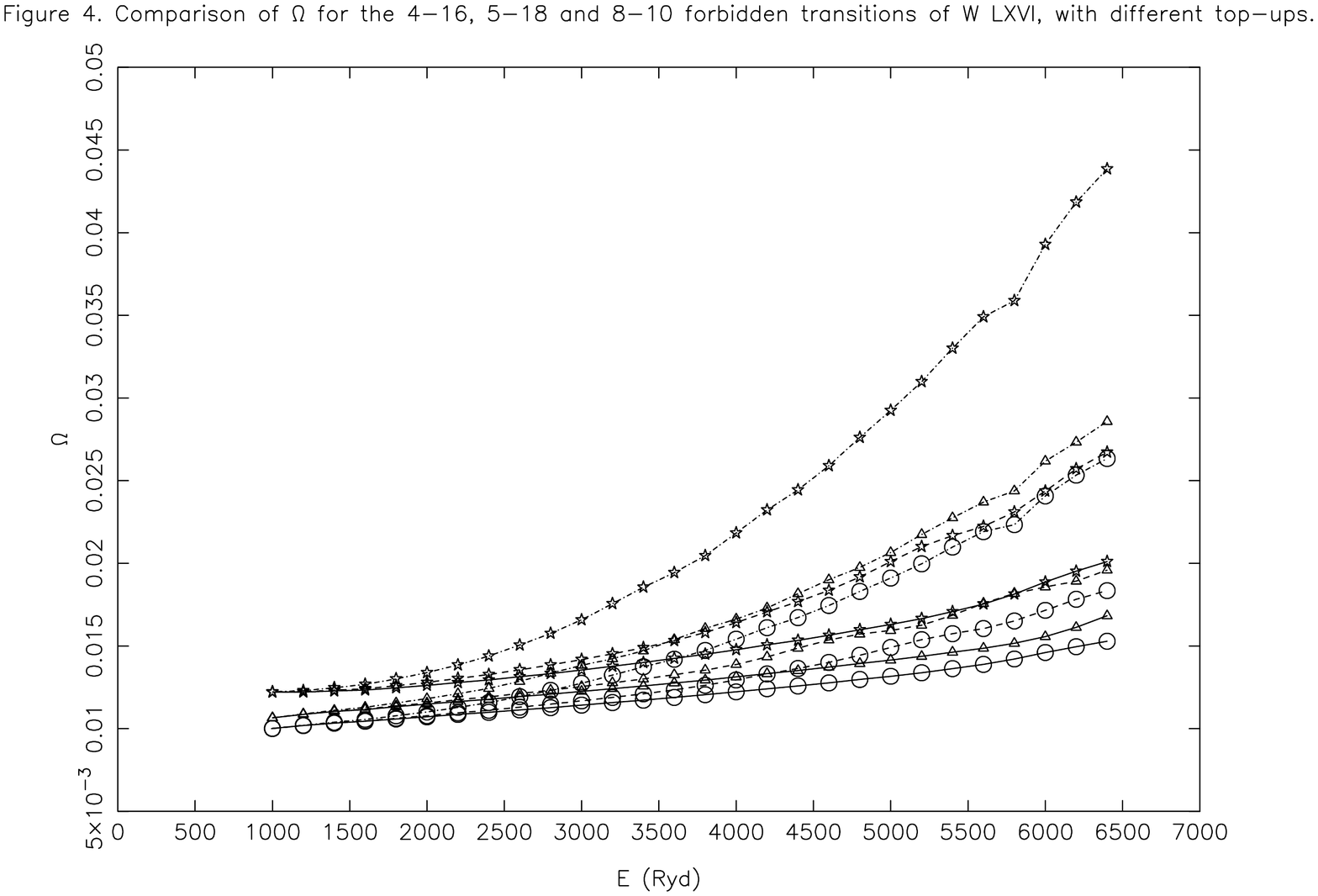}
\caption{Comparison of collision strengths for the 4--16 (2s$^2$2p$^4$3s~$^4$P$_{5/2}$ -- 2s$^2$2p$^4$3d~$^4$D$_{5/2}$: circles), 5--18 (2s$^2$2p$^4$3s~$^2$P$_{3/2}$ -- 2s$^2$2p$^4$3d~$^2$F$_{7/2}$: triangles) and 8--10 (2s$^2$2p$^4$3p~$^2$D$^o_{5/2}$ -- 2s$^2$2p$^4$3p~$^4$P$^o_{5/2}$: stars) transitions of W~LXVI. Continuous curves: with top-up at $J$ = 50, broken curves: at $J$ = 40 and dotted curves: at $J$ = 30.}
\end{figure*}
\end{document}